\documentclass[preprint2]{proto}
\usepackage{times}

\newcommand{\refs}{\par\noindent\hangindent=1pc\hangafter=1}
\begin{document}

\title{\textbf{\LARGE Asteroid Family Physical Properties}}

\author {\textbf{\large Joseph R. Masiero}}
\affil{\textbf{\textit{\small NASA Jet Propulsion Laboratory/Caltech}}}
\author {\textbf{\large Francesca DeMeo}}
\affil{\textbf{\textit{\small Harvard/Smithsonian Center for Astrophysics}}}
\author {\textbf{\large Toshihiro Kasuga}}
\affil{\textbf{\textit{\small Planetary Exploration Research Center, Chiba Institute of Technology}}}
\author {\textbf{\large Alex H. Parker}}
\affil{\textbf{\textit{\small Southwest Research Institute}}}

\begin{abstract}
\begin{list}{ } {\rightmargin 1in}
\baselineskip = 11pt
\parindent=1pc
{\small

An asteroid family is typically formed when a larger parent body
undergoes a catastrophic collisional disruption, and as such family
members are expected to show physical properties that closely trace
the composition and mineralogical evolution of the parent.  Recently a
number of new datasets have been released that probe the physical
properties of a large number of asteroids, many of which are members
of identified families.  We review these data sets and the composite
properties of asteroid families derived from this plethora of new
data. We also discuss the limitations of the current data, and the
open questions in the field. \\~\\~\\~}


\end{list}
\end{abstract}

\section{\textbf{INTRODUCTION}}

Asteroid families provide waypoints along the path of dynamical
evolution of the solar system, as well as laboratories for studying
the massive impacts that were common during terrestrial planet
formation.  Catastrophic disruptions shattered these asteroids,
leaving swarms of bodies behind that evolved dynamically under
gravitational perturbations and the Yarkovsky effect to their
present-day locations, both in the Main Belt and beyond.  The forces
of the family-forming impact and the gravitational reaccumulation of
the collisional products also left imprints on the shapes, sizes,
spins, and densities of the resultant family members (see chapter by
{\it Michel et al.}, in this volume).  By studying the physical properties
of the collisional remnants, we can probe the composition of the
parent asteroids, important source regions of transient populations
like the near-Earth objects, and the physical processes that asteroids
are subjected to on million- and billion-year timescales.

In the thirteen years since the publication of Asteroids III, research
programs and sky surveys have produced physical observations for
nearly two orders of magnitude more asteroids than were previously
available.  The majority of these characterized asteroids are members
of the Main Belt, and approximately one third of all known Main Belt
asteroids (MBAs) with sizes larger than a few kilometers can be
associated with asteroid families.  As such, these data sets
represent a windfall of family physical property information, enabling
new studies of asteroid family formation and evolution.  These data
also provide a feedback mechanism for dynamical analyses of families,
particularly age-dating techniques that rely on simulating the
non-gravitational forces that depend on an asteroid's albedo,
diameter, and density.

In Asteroids III, {\it Zappal\`{a} et al.} (2002) and {\it Cellino et
  al.} (2002) reviewed the physical and spectral properties
(respectively) of asteroid families known at that time.  {\it Zappal\`{a}
et al.} (2002) primarily dealt with asteroid size distributions
inferred from a combination of observed absolute H magnitudes and
albedo assumptions based on the subset of the family members with
well-measured values.  Surveys in the subsequent years have expanded
the number of measured diameters and albedos by nearly two orders of
magnitude, allowing for more accurate analysis of these families.
{\it Cellino et al.} (2002) discussed the spectroscopic properties of the
major asteroid families known at that time.  The principal leap
forward since Asteroids III in the realm of spectroscopy has been the
expansion of spectroscopic characterization to near-infrared (NIR)
wavelengths.  The development of more sensitive instrumentation
covering the $1~\mu$m and $2~\mu$m silicate absorption features and
new observing campaigns to acquire NIR spectra for a large number of
objects have revolutionized studies of asteroid composition and space
weathering.

By greatly expanding the number of family members with measured
physical properties, new investigations of asteroid families can be
undertaken.  Measurements of colors and albedo allow us to identify
outliers in our population lists and search for variations in surface
properties of family members that might indicate heterogeneity of the
parent body or weathering processes.  Diameter measurements let us
build a size frequency distribution and estimate the original parent
body size, both of which are critical to probing the physics of giant
impacts.  Spectra provide detailed mineralogical constraints of family
members, allowing for more sensitive tests of space weathering and
parent heterogeneity, albeit for a smaller sample size, while also
probing the formation environment and allowing for comparisons to
meteorite samples.

In this chapter, we highlight the datasets that have been obtained
since Asteroids III which have greatly expanded our ability to
understand asteroid families.  We discuss their implications for
specific families, and tabulate average photometric, albedo, and
spectroscopic properties for 109 families identified in the chapter by
{\it Nesvorn\'{y} et al.} in this volume.  We also discuss the key questions
that have been answered since Asteroids III, the ones that remain
open, and the new puzzles that have appeared over the last decade.

\bigskip
\section{\textbf{NEW DATA SETS}}
\bigskip

The field of asteroid research has benefited in the last 13 years from
a huge influx of data.  Many of these large sets of asteroid
characterization data (including photometric and thermophysical data)
have been ancillary results of surveys primarily designed to
investigate astrophysical sources beyond the solar system.
Simultaneously, observing programs designed to acquire more
time-intensive data products such as asteroid spectra, photometric
light curves, or polarimetric phase curves have also blossomed.  We
review below the main datasets that have advanced family
characterization in recent years.

\bigskip
\noindent
\textbf{ 2.1 Optical colors from SDSS}
\bigskip

The Sloan Digital Sky Survey (SDSS, {\it York et al.} 2000) produced
one of the first extremely large, homogeneous data sets that contained
information about asteroid surface properties at optical wavelengths.
These data are archived in the SDSS Moving Object Catalog ({\it
  http://www.astro.washington.edu/users/ivezic/sdssmoc/sdssmoc.html})
which currently contains $471,569$ entries of moving objects from
survey scans conducted up to March 2007.  The catalog entries can be
associated with $104,449$ unique known moving objects, however
$\sim250,000$ entries do not have corresponding associations in the
orbital element catalogs implying the potential for a significant
benefit from future data mining efforts.

While the SDSS 5-color photometric system ($u,g,r,i,z$, with central
wavelengths of $0.3543~\mu$m, $0.4770~\mu$m, $0.6231~\mu$m,
$0.7625~\mu$m, $0.9134~\mu$m, respectively) was not designed with
asteroid taxonomy in mind (unlike previous surveys such as the Eight
Color Asteroid Survey, {\it Zellner et al.} 1985a), the sheer size of the
dataset coupled with its extremely well-characterized performance has
made SDSS an immensely valuable asset for defining asteroid families
and exploring their properties.  The near-simultaneous optical color
information obtained during the course of the SDSS survey can be used
to infer the spectroscopic properties of tens of thousands of
asteroids at optical wavelengths.

The Sloan Digital Sky Survey, primarily designed to measure the
redshifts of a very large sample of galaxies, serendipitously observed
many asteroids over the course of its several survey iterations. Under
standard survey operations with $53.9$ second exposures, its
five-color camera was sensitive to stationary sources as faint as
$r\sim22.2$, with similar performance in $u$ and $g$ and somewhat
brighter limits in $i$ and $z$ ($21.3$ and $20.5$, respectively).
To enable the accurate determination of photometric redshifts,
high-precision internal and absolute calibrations were essential. The
care and effort expended on these calibrations carried over into the
dataset of moving object photometry, resulting in the largest
well-calibrated dataset of multi-band asteroid photometry to date.

As of the latest data release, the Moving Object Catalog 4 (MOC4,
{\it Parker et al.} 2008) contains asteroid observations from $519$ survey
observing runs. The automatic flagging and analysis of moving objects
required that they be brighter than magnitude $r=21.5$. The brightest
object in the sample is $r’\sim12.91$, giving the survey a dynamic
range of over 8.5 magnitudes. The large sample size and dynamic range
of this survey make it a powerful tool for studying luminosity
functions of dynamically- or photometrically-selected sub-populations
of the moving objects, of which asteroid families are a natural
example.


Because of the much smaller sample size of asteroids with $u$-band
observations having photometric errors $<0.1~$mag ($44,737$, compared
with $442,743$ with a similar precision in $r$-band), most large-scale
asteroid studies using the SDSS photometry have considered only the
four longer-wavelength bands. These are often further collapsed into
two principle components, the $a^\star$ color ($a^\star = 0.89 \left(g
- r\right) + 0.45 \left(r - i\right) - 0.57$) and the ($i-z$) color
(see {\it Ivezi\'{c} et al.} 2001).

Using the SDSS data set, {\it Ivezi\'{c} et al.} (2002) showed that
asteroid families are easily identified from their optical
colors. They found four primary color classes of families, which
describe one of four characteristic compositions: Vesta, Koronis, Eos,
and Themis.  {\it Jedicke et al.} (2004) and {\it Nesvorn\'{y} et al.}
(2005) expanded upon this to investigate space weathering and find
that older asteroid families have steeper spectral slopes from 0.55 to
1 micron than younger asteroid families.

{\it Szab\'{o} \& Kiss} (2008) used SDSS photometry to constrain the shape
distributions of eight large asteroid families.  By assuming the
observed magnitude differences between different epochs of observation
were uncorrelated, they showed that both older families and families
closer to the Sun tend to have more spherical shape distributions.
This effect is what would be expected for a system where small-scale
cratering collisions redistributed material into gravitational lows,
resulting in a more spherical shape.

{\it Parker et al.} (2008) used the family color relationships found
by {\it Ivezi\'{c} et al.} (2002) as a springboard to refine the
definition of asteroid families beyond solely dynamical
relationships. The ability to spectrophotometrically-refine the sample
of objects linked to each family meant family membership could be
extended further into the background, and allowed the identification
of diffuse family ``halos'' of compositionally-distinct objects ({\it
  Parker et al.} 2008, {\it Bro\u{z} \& Morbidelli} 2013, {\it Carruba
  et al.} 2013). Overlapping but compositionally-distinct families
such as Flora and Baptistina are easily separated with the addition of
color information.

Using the wealth of SDSS data, {\it Carvano et al.} (2010) modeled
established taxonomic classes to define the SDSS color parameters of
each class.  They then used these constraints to perform photometric
taxonomic classifications of SDSS-observed asteroids, including many
large asteroid families.  This technique was later extended by {\it DeMeo
\& Carry} (2013) to better distinguish the boundaries between taxonomic
classes.  This taxonomic strategy can be applied to future photometric
surveys to rapidly classify asteroids as a tool for improving family
associations.  While the resulting taxonomic grouping is of lower
certainty than spectroscopic classification, spectrophotometric
taxonomy offers a fast way to quantify the spectral behavior of large
numbers of asteroids and provide high-quality candidates for
spectroscopic followup.

\bigskip
\noindent
\textbf{ 2.2 Visible and IR Spectroscopic Surveys}
\bigskip

Spectroscopic measurements provide unparalleled sensitivity to the
mineralogical features on the surfaces of asteroids.  As members of
families come from the same progenitor object, surveys of many family
members allow us to probe the composition of that parent body and
search for mineralogical changes indicative of geological processes
during the parent's formation and early evolution.  For families shown
to have homogeneous spectral properties, these surveys can also be
used to identify outlier objects that are compositionally distinct
from the family, especially large objects that can present serious
complications to the analysis of family evolution.

A major focus in the study of families has been the characterization
of the effects of space weathering on the surfaces of atmosphereless
bodies.  Recent work using dynamical simulations to determine the ages
of asteroid families (see the chapter by {\it Nesvorn\'{y} et al.} in
this volume for a discussion of these works) has opened a new avenue
into investigating the potential effects of space weathering on
asteroid surfaces.  Assuming that the family-forming impact completely
refreshed the surfaces of family members, we can look for families
with similar compositions but different ages to search for spectral
changes that would be the hallmark of space weathering.  A number of
researchers have conducted spectroscopic surveys of families to this
end, although the results of these studies have been somewhat
conflicting (e.g. {\it Nesvorn\'{y} et al.}  2005, {\it Willman et
  al.} 2008, {\it Vernazza et al.} 2009, {\it Thomas et al.} 2011,
{\it Thomas et al.} 2012).

Shortly after the publication of Asteroids III, large scale visible
wavelength (0.45-0.9 microns) spectral surveys were published that
were major driver of asteroid compositional studies. {\it Bus \&
  Binzel} (2002) and {\it Lazzaro et al.} (2004) published 1341 and
820 asteroid spectra, respectively, providing a wealth of data for
asteroid studies. Also around this time near-infrared spectrometers
became widely available, such as SpeX on the NASA IRTF ({\it Rayner et
  al.} 2003). The near-infrared provides wavelength coverage that
allows better characterization of silicate features ({\it Burbine \&
  Binzel} 2002, {\it Gaffey et al.} 1993). While no single large
observing campaign has been initiated for main belt asteroids in the
near-infrared, many small programs have been carried out that targeted
specific asteroid families ({\it Sunshine et al.} 2004; {\it
  Mothe-Diniz et al.} 2005; {\it Vernazza et al.} 2006; {\it
  Mothe-Diniz et al.} 2008a; {\it Mothe-Diniz \& Nesvorn\'{y}} 2008b;
{\it Mothe-Diniz \& Nesvorn\'{y}} 2008c; {\it Willman et al.} 2008;
{\it Harris et al.} 2009; {\it Reddy et al.} 2009; {\it de Sanctis et
  al.} 2011; {\it Reddy et al.} 2011; {\it Ziffer et al.} 2011; {\it
  de Leon et al.} 2012).  A compilation of spectral taxonomic
classifications of asteroids is given in the Planetary Data System by
{\it Neese} (2010) and is periodically updated.  We highlight here some of
the specific families that were the subject of spectral
investigations.

{\it Vernazza et al.} (2006) obtained spectra of Karin family members,
finding them to be very similar to OC meteorites.  {\it Willman et
  al.} (2008, 2010) extended this survey in an attempt to constrain
space weathering rates for asteroid surfaces, finding that spectral
slopes are altered on the billion year timescale.  {\it Harris et
al.} (2009) surveyed the Karin family using thermal infrared
spectroscopy, and found albedos lower than expected for fresh asteroid
surfaces, implying space weathering alters albedos on the million year
timescale, in conflict with the {\it Willman et al.} results.

{\it Nathues} (2010) used visible and NIR spectroscopy of 97 Eunomia
family members to study the potential differentiation history of the
family parent.  They show that the majority of family members have
S-type taxonomy and the parent body was likely not fully
differentiated, but may have undergone partial differentiation.  {\it
  Fieber-Beyer et al.} (2011) obtained NIR spectroscopy of 12 Maria
family members.  From these data they associate this family with
mesosiderite-type meteorites, a type iron-rich meteorite thought to
originate from a differentiated parent body.

{\it Ziffer et al.} (2011) obtained NIR spectra of 13 asteroids in the
Themis and Veritas families and find distinct differences in the
spectral behavior of the two families.  They associate both families
with CM2 chondrites, but find no evidence for space weathering of
C-type objects.

{\it Reddy et al.} (2009,2011) performed a spectroscopic survey of the
Baptistina family aimed at constraining the composition of the family
members.  They found a range of compositions represented, but smaller
objects as well as (298) Baptistina itself show clear association with
LL-type ordinary chondrites.  Ordinary chondrite meteorites show two
distinct compositional phases, one at albedos similar to S-complex
asteroids and one that is darker with muted silicate absorption bands.
This darker material has been associated with former surface regolith
which had gases implanted by the solar wind and was later re-lithified
and impact-shocked, resulting in a significantly different
reflectivity with only a nominal difference in composition ({\it Britt
  \& Pieters}, 1991).  Studies of the recent Chelyabinsk meteorite
({\it Kohout et al.} 2014, {\it Reddy et al.} 2014) further confirm
that shock-darkening may play a critical role in the evolution of
asteroid family spectra (cf. {\it Cellino et al.}, 2002).

{\it Licandro et al.} (2012) used the Spitzer space telescope to measure
mid-infrared spectra for eight members of the Themis family, and
determined their albedos and diameters.  They were also able to
constrain the surface thermal inertia and set limits on the surface
composition, showing that family members must have a very low surface
density.  

{\it Jasmim et al.} (2013) performed a spectroscopic survey of objects
classified as Qp in the SDSS spectrophotometric system in the Vesta
family, but were not able to find any significant differences between
these objects and canonical V-type asteroid spectra.  This highlights
the potential ambiguities in spectrophotometric taxonomic
classifications, and the need for further spectroscopic followup of
interesting asteroids to confirm their spectral behavior.

Recently, {\it Vernazza et al.} (2014) mineralogically analyzed spectra of
six S-complex asteroid families and 83 background S-type objects and
compared these results to compositions of various ordinary chondrite
meteorites.  They found a bimodality in their olivine/pyroxene mineral
diagnostic for S-type asteroids which traces the compositional
gradients measured for metamorphosed meteorites with a range of iron
contents.  This bimodality also extended to families, with Koronis,
Agnia, Merxia, and Gefion more closely matching high-iron ordinary
chondrites and Eunomia and Flora matching either low-iron chondrites
from the interiors of bodies or ordinary chondrites from the
near-surface regions showing little metamorphism.  Mineralogical
assessment of near-infrared spectra thus offers a new method of
probing the compositions and metamorphic histories of these S-complex
families.

\bigskip
\noindent
\textbf{ 2.3 Infrared Space Surveys}
\bigskip

Recent improvements in mid-infrared detector technology have spurred
renewed interest in their use for astronomical observations.
Ground-based thermal infrared detectors can obtain data for a small
subset of the brightest targets, however the thermal background
prohibits large-scale surveys of smaller asteroids.  The space
environment, however, is ideal for thermal infrared surveys of the
sky, and two recent satellites have obtained all-sky survey data at
mid-infrared wavelengths of a large number of asteroids.  A more
complete discussion of these surveys is presented in the chapter by
{\it Mainzer et al.} in this volume, and so here we only highlight a few
relevant aspects.

The AKARI space telescope was launched on 21 Feb 2006 and surveyed the
sky at two thermal infrared wavelengths from 6 May 2006 until the
telescope-cooling liquid helium was exhausted on 28 August 2007,
covering over $96\%$ of the sky ({\it Murakami et al.} 2007, {\it Ishihara et
al.} 2010).  The ``Asteroid Catalog Using AKARI'' (AcuA, {\it Usui et
al.} 2011, {\it http://darts.jaxa.jp/ir/akari/catalogue/AcuA.html})
database summarizes the asteroid survey data.  Using the Standard
Thermal Model ({\it Lebofsky et al.} 1986, {\it Lebofsky \& Spencer} 1989),
diameters and visible albedos were derived for $5120$ asteroids.

AKARI also performed spectroscopic observations of 70 asteroids during
the warm mission phase after the cryogen was exhausted, many of which
are the largest members of asteroid families.  These observations
provide unique spectroscopic data covering the wavelength range from
$2-5~\mu$m ({\it Usui et al.}, 2011; {\it Usui et al.}, 2015, in
preparation).  {\it Kasuga et al.} (2012) present physical properties
of Cybele family members determined from this dataset, and discuss the
taxonomic composition of this family by combining spectrally derived
taxonomy with infrared photometry.

The Near-Earth Object Wide-field Infrared Survey Explorer (NEOWISE)
mission launched on 14 December 2009 and surveyed the entire sky at
four infrared wavelengths over the course of its cryogenic
mission. The survey continued during its post-cryogenic and
reactivated missions at the two shortest wavelengths.  The NEOWISE
data catalogs and images are publicly archived at the NASA/IPAC
Infrared Science Archive \newline ({\it
  http://irsa.ipac.caltech.edu/Missions/wise.html}) and contain
thermal infrared measurements of over 150,000 asteroids ({\it Mainzer
  et al.}, 2011, 2014).  Many of these were identified as members of
asteroid families, allowing for the determination of diameter and
albedo distributions for an unprecedented number of families.

{\it Masiero et al.} (2011) presented albedos and diameters for over 120,000
Main Belt asteroids, including over 32,000 members of 46 families from
{\it Nesvorn\'{y}} (2012).  They showed that the albedo distribution within a
family is usually very narrow, but some families have significant
contamination from background objects or show a mixing of multiple
families that overlap in proper-element space.  {\it Masiero et al.} (2013)
split the Main Belt into distinct orbital and albedo components and
showed that overlapping families such as Polana and Hertha (aka Nysa)
can be distinguished easily.

{\it Masiero et al.} (2013) used the NEOWISE diameters to measure the
size-frequency distribution (SFD) for 76 asteroid families that they
identify in their data. The SFD can be computed as $N\propto D^\alpha$
where $N$ is the number of objects with diameter greater that $D$ and
$\alpha$ is the SFD slope.  They found that larger families tended to
show cumulative SFD slopes that converge toward the value of
$\alpha=-2.5$ as expected for a collisionally equilibrated population
({\it Dohnanyi} 1969), while smaller families have a wider dispersion.  The
family associated with (31) Euphrosyne has an anomalously steep SFD
slope, which {\it Carruba et al.} (2014) explain as the result of a
dynamical draining of the largest family fragments by the $\nu_6$
resonance, which runs through the center of this family in semimajor
axis-inclination space.

{\it Ali-Lagoa et al.} (2013) used the NEOWISE albedo measurements to
show that the Pallas family has visible and infrared albedos that are
distinct from the majority of B-type asteroids.  {\it Masiero et al.}
(2014) presented the $3.4~\mu$m albedo distributions of 13 asteroid
families.  They found that the asteroid families form three distinct
groupings of albedos at this wavelength.  Additionally, the Eos family
has unique near-infrared reflectance properties, which likely traces a
mineralogy not seen elsewhere in the Main Belt.

\bigskip
\noindent
\textbf{ 2.4 Asteroid Light Curves and Phase Curves}
\bigskip

Over the last decade, a number of groups have carried out surveys of
asteroid light curves, many focusing on the properties of specific
families.  Light curve analysis from a single epoch can provide
rotation periods and constraints on amplitude, while multi-epoch
observations can allow for rotational pole determination, shape model
fitting, searches for binarity, and detection of non-principal axis
rotation.  The binary fraction is an important test of formation
mechanisms, while pole determination for a significant number of family
members allows for constraints on YORP evolution of family spin
states.

One of the key results from these lightcurve studies has been the
confirmation of the YORP effect on spin poles of objects with similar
ages ({\it Vokrouhlick\'{y} et al.} 2003).  The YORP mechanism results from
asymmetries in the thermal re-emission of absorbed light, creating a
torque on the asteroid.  This can alter the rotation rate of these
objects, but also is predicted to rapidly reorient the spin poles of
most small objects to be perpendicular to the body's orbital plane.
This will result in spin poles clustered near $0^\circ$ and
$180^\circ$ from the orbital pole, and a spin rate distribution
differing from the Maxwellian distribution that is expected for a
collisionally equilibrated system.  {\it Slivan et al.} (2003, 2008, 2009)
measured light curves for 30 members of the Koronis family, showing a
strongly anisotropic distribution of spin poles and a non-Maxwellian
rotation rate distribution, both consistent with YORP-induced rotation
changes.

Surveys of other families have come to similar conclusions: {\it Warner et
al.} (2009) surveyed the Hungaria family asteroids and determined
rotation properties for over 100 of these objects.  They find a
significant excess of slow-rotators, which they attribute to
YORP-induced spin down, as well as a binarity fraction of $15\%$,
similar to the NEA population.  {\it Kryszczy\'{n}ska et al.} (2012)
surveyed 55 objects in the Flora family to determine their rotation
periods and poles.  They found that the Flora family rotation rates
are non-Maxwellian, as is expected from long-term YORP evolution.  {\it Kim
et al.} (2014) determined rotation properties for 57 Maria family
members, finding an excess of prograde rotation states consistent with
the predictions of YORP and Yarkovsky evolution resulting in
retrograde rotators moving into the 3:1 Jupiter resonance and being
removed from the family.

{\it Hanu\u{s} et al.} (2011, 2013) used shape models derived from
light curve inversion to study the spin pole positions of ten asteroid
families.  They found a clear distinction in spin pole direction,
where objects at semimajor axes smaller than the largest remnant have
retrograde rotations while objects at larger semimajor axes have
prograde rotations, as is expected from a system evolving under the
Yarkovsky effect.

Using the large amount of photometric data available from the Minor
Planet Center and the AstOrb database kept by Lowell Observatory,
{\it Oszkiewicz et al.} (2011, 2012) rederived the photometric phase
functions of all known asteroids, applying the improved $H,G_1,G_2$
and $H,G_{12}$ phase equations.  They find that asteroid families show
statistically similar phase slope parameters and that the major
taxonomic complexes can be statistically differentiated.  However,
these distributions overlap, significantly limiting the use of phase
curve parameters for direct family or taxonomy identification.  {\it Bowell
et al.} (2014) searched the Lowell data for statistical anisotropies in
the ecliptic longitude distribution of rotation poles for asteroid
family members.  They show that the four largest groups considered
(the Flora, Vesta, and Koronis families, and the Nysa-Polana complex)
have anisotropic spin longitude distributions showing an excess in the
$30-110^\circ$ longitude range and a dearth in the $120-160^\circ$
longitude range, consistent with the distribution seen for the Main
Belt as a whole.  The authors suggest this may be due to pole
reorientation by the YORP effect, but that extensive modeling is
required as well as consideration of potential selection biases.

\bigskip
\noindent
\textbf{ 2.5 Polarimetric Surveys of Families}
\bigskip

Measurement of the polarization of light scattered off the surface of
an atmosphereless body as a function of phase angle can be used as an
independent constraint on the object's albedo ({\it Cellino et al.}, 1999).
Although the magnitude range accessible to polarimetric observations is
comparable to that of NIR spectroscopy, multiple nights are required
for each object to constrain the phase behavior and there are fewer
available imaging polarimeters that are well-calibrated than NIR
spectrometers.  As such, there are fewer asteroids with measured
polarimetric properties, however polarimetric studies provide unique
data on asteroid surface properties that can be important for
disentangling ambiguities in other characterization surveys.

{\it Cellino et al.} (2010) polarimetrically surveyed the Karin and Koronis
families as an independent test of the effect of space weathering on
asteroid albedo.  They find no significant differences between the
albedos of Karin and Koronis, implying that space weathering must act
on timescales shorter than the $\sim 6~$million year age of the Karin
family.

{\it Cellino et al.} (2014) conducted a survey of the linear
polarization phase curves of Watsonia asteroid family members,
searching for analogs to the unusual polarization of (234) Barbara.
Barbara and the related ``Barbarian'' objects have unusual
polarimetric properties indicative of unique surface mineralogy and
are taxonomically identified as Ld-type objects from spectra.  They
found that seven of the nine objects surveyed show indications of a
surface rich in spinel, one of the oldest minerals in the solar system
({\it Burbine et al.}, 1992).  This implies that the Watsonia parent
object has undergone little-to-no mineralogical processing since the
formation of the first solids in the protosolar disk ({\it Sunshine et
  al.}, 2008), and may have been one of the oldest unprocessed bodies
in the Solar system.

\bigskip
\section{\textbf{COLOR, SPECTRA, ALBEDO, AND SIZE DISTRIBUTIONS OF FAMILIES}}
\bigskip

With the wealth of asteroid physical observations now available, we
can determine the average characteristic properties for families.  As
the literature contains many different collections of family lists,
here we focus on the consolidated list presented in the chapter by
{\it Nesvorn\'{y} et al.} in this volume.  This improves our sensitivity to
subtle physical differences between families, especially the biggest
families where the large sample size can greatly reduce the scatter in
measured properties among individual members.  We can also use these
data to reject outlier objects from family lists, particularly when
the family of interest is of distinct composition from other nearby
families or the majority of background objects.  Here we discuss some
of the recent applications of the new physical property data that has
become available.

\bigskip
\noindent
\textbf{ 3.1 Homogeneity of Families and Use of Physical Properties to Distinguish Outliers}
\bigskip

Family-forming collisions are expected to liberate material from a
large fraction of the parent body's volume.  Many of the resulting
family members are likely to be accumulations of smaller ejecta,
potentially from a range of lithologies within the parent (see the
chapter by {\it Michel et al.} in this volume).  For a heterogeneous
parent body, we would expect to see a range of colors, spectra, and
albedos among the resultant family members, while a homogeneous parent
would instead produce a family with very narrow distributions of these
properties.  While it is possible that the family formed from an
impact on a homogeneous parent by an impactor of a different
composition could show heterogeneity, statistically the mass of any
probable impactor impactor will be only a small fraction of the mass
of either the parent or the ejecta, and so would be unlikely to be a
significant contaminant to globally averaged properties obtained from
remote sensing.  Sub-families resulting from second-generation impacts
could also show subtle property differences compared to the original
family due to the reset of any space weathering processes, however
this would be most obvious only for very young families (e.g. Karin).

Early surveys of asteroid family spectra and colors indicated that
with the exception of a handful of outliers (e.g. the Nysa-Polana
complex) families had narrow distributions of physical properties,
indicative of homogeneous parent bodies (see {\it Cellino et al.},
2002 and references therein).  This enabled the extrapolation of
measured physical properties for a handful of family members to the
entire family.  As many of the largest remnants of family formation
had been studied for decades, this became a boon to family research.

However, with the availability of physical property data for large
numbers of family members, we can now use this data to remove
interloper objects from family lists when they have a different color,
albedo, or spectrum from the bulk of the family.  Refining family
lists in this way is particularly important for improving the accuracy
of age-dating techniques.  In the past few years, a number of research
groups have been actively using these data to this end, in an effort
to better understand the formation and evolution of families and the
parent bodies they came from (e.g. {\it Novakovi\'{c} et al.} 2011;
{\it Bro\u{z} et al.} 2013; {\it Masiero et al.} 2013; {\it Carruba}
2013; {\it Walsh et al.} 2013; {\it Carruba et al.} 2013; {\it
  Hanu\u{s} et al.} 2013; {\it Milani et al.} 2014).

\bigskip
\noindent
\textbf{ 3.2 Combined Properties of Individual Families}
\bigskip

\begin{figure*}
\epsscale{2.2} \plotone{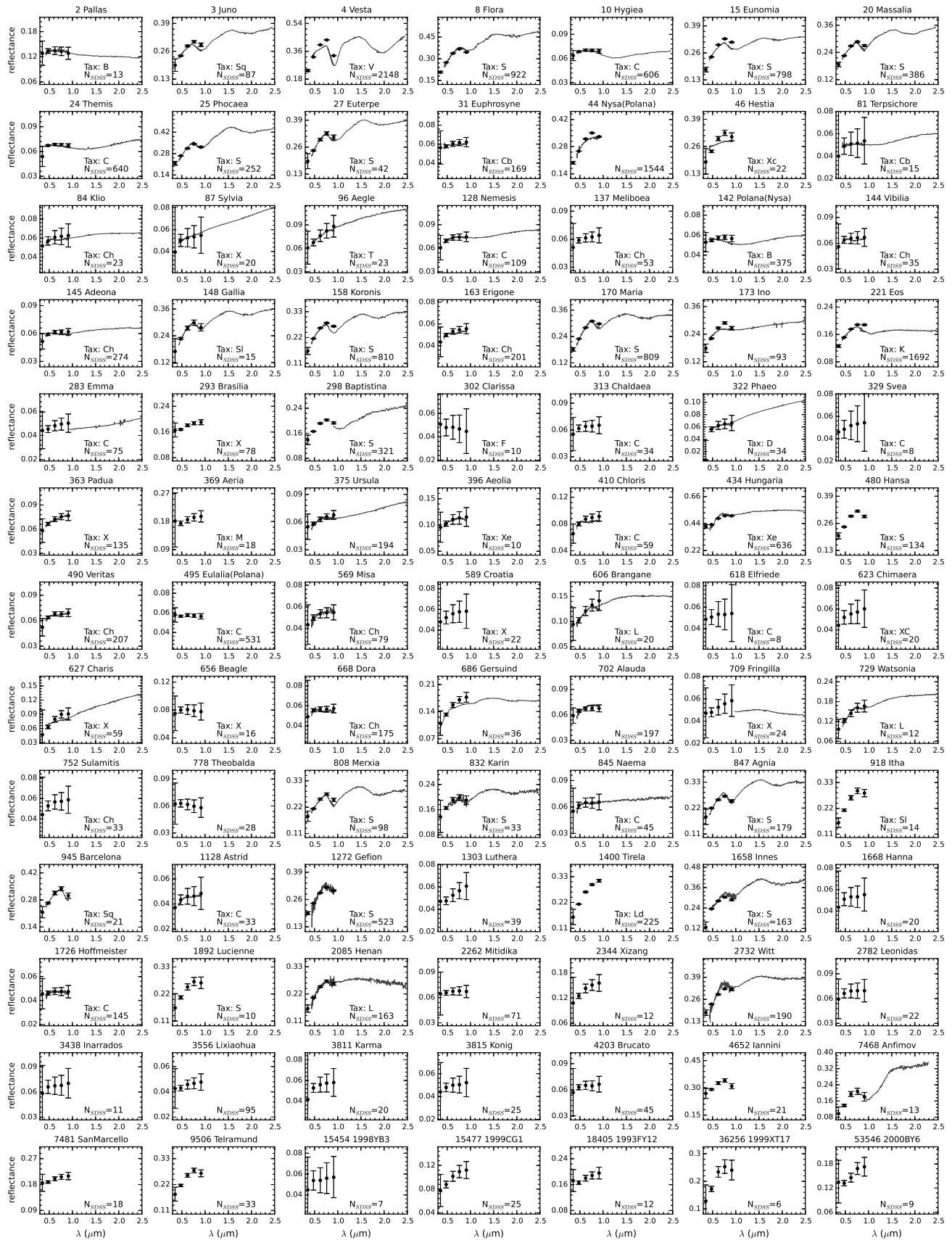}
\caption{\small Average solar-corrected SDSS colors (points) and
  sample optical/NIR spectra (from SMASS) for all asteroid families
  listed in Table~\ref{tbl.fams} with sufficient data.  Plots are
  scaled such that the interpolated reflectance at $0.55~\mu$m equals
  the average visible geometric albedo ($p_V$) for the family from all
  infrared surveys.  The ``N'' in the bottom right of each plot
  indicates the number of SDSS observations used, and taxonomic class
  is given when available.  Note the reflectance scale in each plot is
  different.}
\label{fig.spec}
\end{figure*}

Using the various data sets presented above, we provide in Table
\ref{tbl.fams} the average SDSS colors, average optical and NIR
albedos, majority taxonomic type, and SFD slopes (as well as the
diameter range that the SFD was fit to) for the families given in the
chapter by {\it Nesvorn\'{y} et al.} in this volume.  We also present a
visualization of the average albedo, SDSS colors and spectra for
families with sufficient data in Figure \ref{fig.spec}.  While mean
values are often useful for extrapolating properties of family members
that were not observed, or for comparing families to each other or
other ground-truth data, there are important caveats that cannot be
ignored.  Sample size considerations and the uncertainty on individual
measurements are the most critical of these caveats: average values
based on a small number of objects or even a single object should be
considered potentially spurious, and treated as such.  We include
those values here for completeness.

Visible albedo mean values were derived by fitting a Gaussian to the
distribution of all measured $\log{p_V}$ values, following the
technique {\it Masiero et al.} (2011) applied to the NEOWISE data, and
the quoted error gives the width of the observed distribution.  The
uncertainty on the mean value, when comparing mean albedos of various
families, will be the measured Gaussian width over the square root of
the sample size; thus for large families the mean albedos can be known
quite accurately, even if individuals have large uncertainties. Family
NIR albedo values are a median of the $3.4~\mu$m albedos given in
{\it Masiero et al.} (2014).

Size frequency distributions (SFDs) were computed for all families
with more than $100$ measured diameters from all infrared surveys and
were found by fitting a power law to the cumulative size distribution,
using only those bins with more than $5$ objects (to reduce the
influence of the large remnants) and less than half the total sample
(to minimize the effects of incomplete diameter sampling at small
sizes).  Tabulated SDSS colors were calculated by performing an
error-weighted mean of all observations having SNR$>10$ in the
$g$,$r$,$i$, and $z$ bands used to calculate $a^\star$ and $i-z$.
Family taxonomy is given only for cases where a majority of family
members with taxonomic classes had the same class.  If no subclass
(e.g. Ch, Sq, Ld) had a majority, but a majority of members were in
the same complex, that complex is listed (e.g. C, S, L).  For each of
these parameters we include the number of objects used to compute the
listed value.

Although we implicitly assume that the variation in observed
properties is a result of statistical uncertainties and that all
family members should have the same physical properties, this is not
necessarily true.  Outlier objects may be statistical flukes, or
background contamination, but they just as easily may be interesting
pieces of the parent body warranting further study, a determination
that cannot be made or captured here.  Additionally, a single mean
value ignores any size-dependent effects in the data, either real or
imposed by detection or sample-selection biases.

Only a proper debiasing of each survey accounting for observing
geometry, detector sensitivity, and detection efficiency can determine
the true value for each parameter.  This is particularly important for
cases of mismatched sensitivities, such as the SDSS $r$ vs $z$
magnitudes needed for colors or the NEOWISE W3 flux vs ground-based
$H_V$ magnitude needed for albedo determination; only objects seen in
both datasets will have a measured value which strongly biases the
outcome as a function of apparent brightness.  Thus Table
\ref{tbl.fams} is meant as a overarching guide, but caution is
mandated for any interpretation of values or trends.

A number of low albedo families can only be identified when the low
albedo population of the Main Belt is considered independently.  This
is due to the bias toward discovery of high-albedo objects by
ground-based visible-light surveys.  As the population of high albedo
asteroids is probed to smaller sizes (and thus larger numbers) these
families will overwhelm traditional HCM techniques, particularly the
calculation of the quasi-random level needed to assess the reality of
a given family, making low albedo families harder to identify.
{\it Masiero et al.} (2013) circumvented this by considering each albedo
component separately, and by restricting their sample to objects
detected in the thermal infrared, which is albedo-independent (see the
chapter by {\it Mainzer et al.} in this volume for further discussion).  As
the sample of known asteroids continues to increase, development of
new techniques for the identification of asteroid families from the
background population (e.g. {\it Milani et al.} 2014) will increase in
importance.

Below we discuss individual families that merit specific mention based
on recent research.  We include the PDS ID number from {\it
  Nesvorn\'{y} et al.} (2012) both below and in Table \ref{tbl.fams}
for easier association with other work and with the family dynamical
properties given in the chapter by {\it Nesvorn\'{y} et al.} in this
volume.

\bigskip
\noindent
\textit{ 3.2.1 Hungaria}
\bigskip

The Hungaria asteroid family (PDS ID 003) occupies a region of space
interior to the rest of the Main Belt and with an orbital inclination
above the $\nu_6$ secular resonance.  This region is an island of
stability between the major resonances that dominate this area of the
solar system, and likely samples a unique region of the protoplanetary
disk ({\it Bottke et al.} 2012).  {\it Warner et al.} (2009) analyzed
light curves of $129$ Hungaria asteroids, finding a significant excess
of very slow rotating bodies.  They also showed that the binary
fraction of this population is $\sim15\%$, comparable to the fraction
seen in the NEO population.

The albedos for Hungaria family members derived from the NEOWISE data
by {\it Masiero et al.} (2011) have values significantly larger than
$p_V>0.5$, however this is an artifact of bad absolute magnitude fits
for these objects in orbital element databases, which when corrected
bring the best-fit albedo values to the range of $0.4<p_V<0.5$ ({\it
  B. Warner}, private communication).  Spectroscopic and SDSS color
studies of the Hungaria family show the classification to be X-type,
which when combined with the very high albedos translates to an E-type
classification ({\it Carvano et al.} 2001; {\it Assandri \&
  Gil-Hutton}, 2008; {\it Warner et al.} 2009) Polarimetric
observations by {\it Gil-Hutton et al.} (2007) of the overall Hungaria
region indicate inconsistencies in the polarimetric behavior of
asteroids in and around the Hungaria family.

\bigskip
\noindent
\textit{ 3.2.2 Flora}
\bigskip

The Flora family (PDS ID 401) is a large S-type family residing in the
inner Main Belt.  The largest remnant, (8) Flora, has an orbit just
exterior to the $\nu_6$ resonance, and only the half of the family at
larger semimajor axes is seen today.  The $\nu_6$ resonance is
particularly good at implanting asteroids into the NEO population
({\it Bottke et al.} 2000), meaning that Flora family members are
likely well-represented in the NEO population and meteorite
collections.  Recent spectroscopic observations of Flora family
members have been combined with analyses of meteorite samples to link
the LL chondrite meteorites to the Flora family ({\it Vernazza et al.}
2008; {\it de Leon et al.} 2010; {\it Dunn et al.} 2013).  This
provides an important ground-truth analog for interpreting physical
properties of S-type objects in the Main Belt and near-Earth
populations.  We note that in contrast to previous analyses, {\it
  Milani et al.} (2014) did not identify a family associated with
Flora, instead finding that candidate member asteroids merged with the
Vesta and Massalia families, or potentially are part of their newly
identified Levin family.  However, the Vesta family has a distinct
$i-z$ color that is not shown by Flora family members (see Table 1),
making these populations easily distinguished by their photometric
properties.  While the Flora and Massalia populations overall have
properties that are consistent within uncertainties, the difference
between the mean albedos of these two populations suggests they in
fact are different populations.

\clearpage
\bigskip
\noindent
\textit{ 3.2.3 Baptistina}
\bigskip

Over the past decade, the Baptistina family (PDS ID 403) has been the
focus of a number of investigations, leading to controversy over the
physical characteristics of these asteroids.  Initial investigations
assumed the family had characteristics similar to C-type asteroids
(e.g. low albedo) based on spectra of the largest member (298)
Baptistina ({\it Bottke et al.}, 2007).  This was used in numerical
simulations to show that Baptistina was a probable source of the K/T
impactor.  Further spectral investigations of a 16 large family
members found compositions more analogous to LL chondrites, ruling out
a C-type association ({\it Reddy et al.}, 2011).  However, a major
difficulty in studying this family is the significant overlap in
orbital element-space with the much larger Flora family (or with the
Levin family, according to {\it Milani et al.} (2014)), making it difficult
to ensure that the spectral studies were probing Baptistina and not
Flora.  Using albedos to separate these families finds a mean albedo
of $p_V=0.16$, which is not consistent with either C-type or LL
chondrite compositions ({\it Masiero et al.} 2013).  Recent analysis of the
Chelyabinsk meteorite samples showed that shock darkening of
chondritic material could produce an albedo consistent with the
Baptistina family without altering the composition ({\it Reddy et al.} 2014)
offering a potential solution to the seemingly contradictory
information about this family, but future work will be necessary to
confirm this hypothesis.

\bigskip
\noindent
\textit{ 3.2.4 ``Nysa-Polana''}
\bigskip

In the years leading up to Asteroids III it had become increasingly
clear that Nysa-Polana, interpreted as a single entity, was likely to
be a short-lived phenomenon.  {\it Cellino et al.} (2001) presented a
spectroscopic study of 22 asteroids associated with the group, and
found that the group was best understood as two compositionally
distinct families partly overlapping in orbital element space, one
associated with the F class in the Tholen taxonomy and one with the S
class, while (44) Nysa was compositionally distinct and potentially
not associated with either family.

{\it Masiero et al.} (2011) showed that the albedo distribution of the
$\sim3000$ asteroids identified as part of the Nysa-Polana complex
were strongly bimodal, unlike the majority of families.  Although
overlapping in semimajor axis-inclination space, the high- and
low-albedo components occupy distinct regions of semimajor
axis-eccentricity space, supporting the theory that they are two
distinct populations coincidentally overlapping as opposed to a single
parent body that was composed of two distinct mineralogies.  

Using albedo as a discriminant, {\it Masiero et al.} (2013) were able to
uniquely identify two separate families, a low-albedo one associated
with (142) Polana, and a high-albedo one with (135) Hertha (PDS ID
405), while (44) Nysa no longer linked to either family.  In Table
\ref{tbl.fams} we continue to refer to the high-albedo family as
``Nysa'' despite the evidence to the contrary, for consistency with
other literature.  {\it Walsh et al.} (2013) expanded on this, and used the
family albedos to reject objects with S-type physical properties and
focus on the low-albedo component of the Nysa-Polana group.  Using
dynamical constraints, they were able to further sub-divide the
low-albedo component of this group into two distinct families, which
they identify as the Eulalia family and the ``new Polana'' family.
They estimate ages for each of these families of $0.9-1.5~$billion
years and $>2~$billion years, respectively.  Conversely, {\it Milani et
al.} (2014) identify the whole complex as associated with Hertha and
split this region into two components by combining dynamics and
physical properties, which they identify as the Polana and Burdett
families (low and high albedo, respectively).

\bigskip
\noindent
\textit{ 3.2.5 Vesta}
\bigskip

The Vesta asteroid family (PDS ID 401) has historically been one of
the most well-studied families, due to its clear association with one
of the largest known asteroids, the high albedos and locations in the
inner Main Belt making members favorable for ground-based
observations, and association with the HED meteorites leading to the
interpretation of (4) Vesta as a differentiated parent body ({\it
  McCord et al.}, 1970; {\it Zappala et al.}, 1990; {\it Binzel \& Xu}
1993; {\it Consolmagno \& Drake}, 1977; {\it Moskovitz et al.}, 2010;
{\it Mayne et al.}, 2011).  With the recent visit of the Dawn
spacecraft to (4) Vesta (see the chapter by {\it Russell et al.} in
this volume) ground-truth data can be compared to remote-sensing
observations of family members.  Additionally, constraints on the ages
of the major impact basins of $1.0 \pm 0.2~$billion years for
Rheasilvia and $2.1\pm0.2~$billion years for Veneneia ({\it Schenk et
  al.} 2012) set strong constraints on the possible age of Vesta
family members.  {\it Milani et al.} (2014) find that the Vesta family
splits into two subgroups in their analysis, which they attribute to
these two events.

The Vesta family has a unique mineralogical composition in the Main
Belt, making it easily distinguishable in color-, albedo-, or
spectral-space.  In particular, members stand out from all other
asteroids in terms of their high albedo, low $i-z$ color, and deep
$1~\mu$m and $2~\mu$m absorption bands.  This has prompted searches
for objects with similar properties at more distant locations in the
Main Belt (e.g. {\it Moskovitz et al.} 2008; {\it Duffard \& Roig},
2009; {\it Moskovitz et al.}, 2010; {\it Solontoi et al.} 2012).
These bodies could only have evolved from the Vesta family via a
low-probability series of secular resonances ({\it Carruba et al.}
2005, {\it Roig et al.} 2008a), and may be indicative of other parent
bodies that were differentiated.  To date only a few candidate objects
from these searches have been confirmed to be V-type, indicating that
the collisional remnants of the other differentiated objects that must
have formed in the early solar system has likely been dynamically
erased.

\bigskip
\noindent
\textit{ 3.2.6 Eunomia}
\bigskip

The Eunomia family (PDS ID 502) is an old, S-type family located in
the middle Main Belt.  Spectral evidence presented by {\it Lazzaro et al.}
(1999) indicates that the Eunomia family may have originated from a
partially differentiated parent body. {\it Natheus et al.} (2005) and
{\it Nathues} (2010) investigated the physical properties of the Eunomia
largest remnant and 97 smaller family members as a probe of the
composition and differentiation history of the original parent body.
They found that the largest remnant shows two hemispheres with
slightly different compositions that support an interpretation of the
impact causing significant crust-loss and some mantle-loss on a
partially-differentiated core.  {\it Milani et al.} (2014) found two
subfamilies within Eunomia, which they attribute to separate cratering
events.

\bigskip
\noindent
\textit{ 3.2.7 Eos}
\bigskip

The Eos family (PDS ID 606) represents the primary reservoir of K-type
asteroids in the Main Belt, and can easily be identified by their
$3.4~\mu$m albedo ({\it Masiero et al.} 2014).  This spectral class
has been proposed as the asteroidal analog of the CO and CV
carbonaceous chondrite meteorites ({\it Bell et al.} 1988; {\it
  Doressoundiram et al.} 1998; {\it Clark et al.} 2009).  This would
imply that the Eos family is one of the best sampled collisional
families in our meteorite collection, and would mean that many of
these samples do not probe the C-complex asteroids as had frequently
been assumed.

{\it Moth\'{e}-Diniz \& Carvano} (2005) compared spectra of (221) Eos
with meteorite samples and inferred that the Eos parent body likely
underwent partial differentiation.  {\it Moth\'{e}-Diniz et al.}
(2008a) extended this work to 30 Eos family members and found mineral
compositions consistent with forsteritic olivine consistent with a
history of differentiation or a composition similar to CK-type
meteorites.  However, {\it Masiero et al.} (2014) instead interpret
the surface properties as consistent with shock-darkening of silicates
(cf. {\it Britt \& Pieters} 1994, {\it Reddy et al.} 2014).  Future
work will enable us to resolve the ambiguity in the composition of
these objects.  The Eos family represents one of the key fronts of
advancing our understanding of family formation that has been opened
by our wealth of new data.

\bigskip
\noindent
\textit{ 3.2.8 Themis}
\bigskip

The Themis family (PDS ID 602) is one of the largest low-albedo
families in the Main Belt.  The majority of Themis family members are
classified as C-complex bodies ({\it Mothe-Diniz et al.} 2005; {\it
  Ziffer et al.} 2011).  Spectral surveys have found variations in the
spectral slope among Themis members, ranging from neutral to
moderately red ({\it Ziffer et al.} 2011; {\it de Leon et al.} 2012).
As the first asteroid discovered to show cometary activity (133P
Elst-Pizzaro) is dynamically associated with Themis, {\it Hsieh \&
  Jewitt} (2006) searched 150 other Themis family members for signs of
cometary activity, discovering one additional object: (118401) 1999
RE$_{70}$.  The periodic nature of this activity points to volatile
sublimation as the probable cause, as opposed to collisions or YORP
spin up (see the chapter by {\it Jewitt et al.} in this volume)
meaning that the Themis family members, and (24) Themis itself, likely
harbor subsurface ice.  Further, {\it Rivkin \& Emery} (2010) and {\it
  Campins et al.} (2010) reported evidence of water ice features on
the surface of (24) Themis.  If this icy material was primordial to
the Themis parent and not implanted by a later impact, this would set
constraints on where the Themis parent formed relative to the ``snow
line'' in the protosolar nebula.  These objects are also likely to be
a new reservoir of water, and may have contributed to the volatile
content of the early Earth.

\bigskip
\noindent
\textit{ 3.2.9 Sylvia}
\bigskip

The Sylvia family (PDS ID 603) in the Cybele region, resides at the
outer edge of the Main Belt ($3.27<a\le3.70~$AU, {\it Zellner et al.},
1985b), beyond the 2:1 Jupiter mean motion resonance.  These
asteroids, along with the objects in the Hilda and Jupiter Trojan
populations, likely have limited contamination by materials from the
inner solar system, and represent a pristine view of the materials
present near Jupiter at the end of planetary migration.  The Nice
model proposed that these asteroids are trans-Neptunian objects (TNOs)
that were scattered inward during the chaotic phase of planetary
evolution ({\it Levison et al.}, 2009), however the Cybeles show different
spectral characteristics from the Hildas and Trojans, and thus may
represent the material native to this region of the solar system.

{\it Kasuga et al.} (2012) studied the size- and albedo-distributions
of the Sylvia asteroid family to better understand the history of
these bodies.  They found that the largest Cybeles ($D>80~$km) are
predominantly C- or P-types, and the best-fit power law to the size
distribution is consistent with a catastrophic collision.  However the
estimated mass and size of the parent body lead to collisional
timescales larger than the age of the solar system, assuming an
equilibrium collisional cascade.  These are comparable to the
timescales for the Hildas, although they find that the Trojan
population is consistent with collisional origin.  Numerical
simulations of the collisional formation and dynamical evolution of
the Cybeles will allow for a more detailed study of the history of
this population.

\bigskip
\noindent
\textbf{ 3.3 Relationship between albedo and color}
\bigskip

\begin{figure*}
\epsscale{2} \plotone{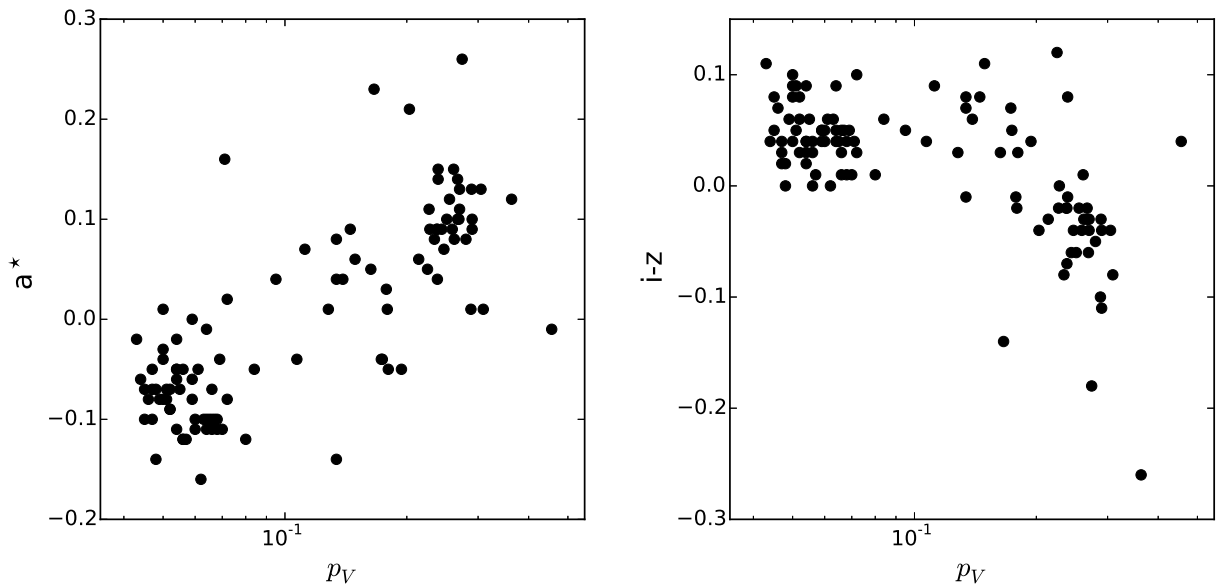}
\caption{\small Average SDSS $a^\star$ and $i-z$ colors compared with optical albedos for asteroid families, as given in Table \ref{tbl.fams}}
\label{fig.compare}
\end{figure*}

Almost every major taxonomic class of asteroid is represented in the
list of asteroid families.  Using the physical property data described
above, we can investigate relationships between the averaged physical
properties for each family.  By nature of the large sample sizes, the
SDSS colors and optical albedos are the best-determined parameters for
the majority of families.  Figure \ref{fig.compare} shows the
composite $a^\star$ and $i-z$ colors derived for each family from SDSS
compared to the composite optical albedo as determined via thermal
radiometry.  There is a clear linear correlation between the SDSS
$a^\star$ color and the log of the albedo, although with significant
scatter.  The $i-z$ color is approximately flat for low and moderate
albedos, but decreases noticeably for high albedo families.  It is
important to again note that these data are subject to observational
selection and completeness biases, and so these relations should be
interpreted with caution.

\bigskip
\section{\textbf{DISCUSSION}}
\bigskip

\bigskip
\noindent
\textbf{ 4.1 Correlation of Observed Properties with the Primordial Composition of the Main Belt}
\bigskip

When considering the distributions of colors and albedos, asteroid
families fall in one of only a small number of groupings.  The
Hungaria, Vesta, and Eos families have unique properties that
distinguish them from all other families.  Similarly, the Watsonia
family shows unique polarimetric properties.  While the properties of
these families can be used to efficiently identify family members and
mineral analogs, it is more difficult to relate the mineralogy and
history of these families to that of the currently observed Main Belt
as a whole (although they may be good analogs for now-extinct
populations, cf. {\it Bottke et al.} 2012).

Conversely, almost all other families fall into either the S-complex
or the C-complex, both of which can be further subdivided by spectral
taxonomic properties. Given the number of different parent bodies with
these compositions, it has been inferred that these two complexes are
probes of the pristine material from the protoplanetary disk in this
region of the solar system.  However, studies of the evolution of the
giant planets in the early solar system call into question the
specific locations these bodies originated from (e.g. {\it Gomes et
al.} 2005).  One possible scenario, known as the ``Grand Tack'' ({\it Walsh
et al.} 2012), postulates that the protoplanetary core of Jupiter
migrated through the planetesimal disk, evacuating the Main Belt
region of its primordial material and repopulating it with material
from both inward and outward in the disk (see the chapter by
{\it Morbidelli et al.} in this volume).  In this case, the two different
compositional complexes would represent these implanted objects.
Further study of the physical properties of asteroid families and the
other small body populations of the solar system will allow us to test
this theory.

\bigskip
\noindent
\textbf{ 4.2 Properties of Observed Families Contrasted with the Background Population}
\bigskip

Using computed $a^\star$ colors from the SDSS MOC 4, {\it Parker et
al.} (2008) divided asteroid families into blue and red groups, tracing
the C and S taxonomic complexes, respectively.  For both groups, they
found that family membership as a fraction of total population
increased with decreasing brightness from $\sim20\%$ at $H_V=9$ to
$\sim50\%$ at $H_V=11$.  For objects with absolute magnitudes of
$H>11$ the fractions of blue and red asteroids in families diverge,
with blue objects in families making up a smaller fraction of the
total population while red objects stay at the levels seen at brighter
magnitudes, however it is unclear what contribution the survey biases
have to these observed differences.  If this difference in behavior is
indeed a physical effect, it may indicate a difference in internal
structure or collisional processing rates between the two populations.

When exploring the taxonomic distribution of asteroids as a function
of distance from the Sun, families have been problematic because the
large number of homogeneous objects concentrated in small regions of
orbital-element space can skew the results.  Studies of the overall
distributions of physical properties therefore typically use only the
largest member of the family, removing the smaller members (e.g. {\it
  Mothe-Diniz et al.} 2003). Alternatively, the distribution could be
explored by volume or mass, in which case all family members may be
included because their individual volumes or masses contribute to the
whole (e.g. {\it DeMeo \& Carry} 2013).

A comprehensive study of the taxonomic contribution of families to the
population of small asteroids has not yet been undertaken, however
disentangling families from the background is critical to correctly
interpreting an overall picture of the compositional make up of the
asteroid belt and how the asteroid belt and the bodies within it
formed. Distinguishing between families and the background becomes
increasingly difficult at smaller sizes where orbital parameters have
evolved further due to gravitational and non-gravitational
forces. Even the background itself is likely composed of many small
families forming from small parent bodies ({\it Morbidelli et al.}
2003).

\bigskip
\noindent
\textbf{ 4.3 Families as Feeders for the NEO Population}
\bigskip

The Yarkovsky and YORP non-gravitational effects play a critical role
in repopulating the near-Earth objects with small bodies from the Main
Belt ({\it Bottke et al.} 2000).  Primordial asteroids with diameters
of $\sim100~$m should be efficiently mobilized from their formation
locations into a gravitational resonance over the age of the solar
system, limiting the contribution of these objects to the currently
observed NEO population.  However, family formation events act as an
important source of objects in this size regime, injecting many
thousands of small asteroids into the Main Belt with each impact
({\it Durda et al.} 2007).  The complete census of family physical
properties, combined with better estimates for family ages that can
now be made, enable us to trace the history of specific near-Earth
asteroids from their formation in the Main Belt to their present day
orbits.  For the same reasons, recently fallen meteorites are also
good candidates for comparisons to asteroid families, however the
differences between the surface properties we can observe on asteroids
and the atmosphere-selected materials surviving to the ground
complicate this picture.

\bigskip
\noindent
\textbf{ 4.4 Families Beyond The Main Belt}
\bigskip

Although the vast majority of known asteroid families are in the Main
Belt, massive collisions resulting in family formation events are not
unique to this region of the solar system.  While the near-Earth
asteroid population is dynamically young ({\it Gladman et al.} 2000) and
thus these objects have a low probability of undergoing collisional
breakup, the more distant reservoirs that date back to the beginning
of the solar system are expected to undergo the same collisional
processing as the Main Belt, albeit with lower impact velocities.
Searches for young families in both the NEO and Mars Trojan
populations have yielded only a single candidate family cluster
associated with (5261) Eureka ({\it Schunov\'{a} et al.} 2012; {\it Christou}
2013).  Dynamical families have been identified in the Hilda, Jupiter
Trojan, and TNO populations.  The first two populations are trapped in
long-term stable resonant orbits with Jupiter, providing a population
that has suffered far less dispersion than the majority of Main Belt
families.  Similarly, the TNO population is cross-cut by a range of
Neptune resonances, some of which are similarly long-term stable.  The
Yarkovsky and YORP effects are greatly diminished for all three
populations compared to the Main Belt (especially the TNOs) due to the
larger distances to the Sun, further reducing the dispersion of
collisional fragments.

{\it Grav et al.} (2012) present visible and infrared albedos for the
Hilda and Schubart asteroid families found in the 3:2 Jupiter mean
motion resonance.  Using these albedos, they show the Hilda family is
associated with D-type taxonomy, while the Schubart family is
associated with C or P-type taxonomy.  The Hilda population in general
is dominated by C and P-type objects at the largest sizes, but
transitions to primarily D-class at the smallest sizes measured, which
may be indicative of the effect of the Hilda family on the overall
population ({\it Grav et al.} 2012; {\it DeMeo \& Carry}, 2014).
However we note that recent family analysis by {\it Milani et al.}
(2014) showed that the Hilda family was not statistically significant
using their methodology, in contrast to previous work.

The Jupiter Trojans are comprised of the L4 and L5 Lagrange point
swarms which lead and trail Jupiter (respectively) on its
orbit. Multiple families have been identified in each of the swarms
({\it Milani} 1993, {\it Beaug\'{e} \& Roig}, 2001), however there is
some debate as to the significance of these families (cf. {\it
  Bro\u{z} \& Rozehnal}, 2011).  {\it Fornasier et al.} (2007)
combined measured spectra from multiple sources ({\it Fornasier et
  al.} 2004, {\it Dotto et al.} 2006) to characterize Trojan family
members.  In the L5 cloud, members of the Aneas, Anchises, Misenus,
Phereclos, Sarpedon, and Panthoos families were found to have spectra
with moderate-to-high spectral slopes, with most members being
classified as D-type. The background population had a wider range of
slopes and taxonomies from P- to D-type ({\it Fornasier et al.},
2004).

In the L4 cloud members of the Eurybates, 1986 WD and 1986 TS$_6$
families were studied. The 1986 WD and 1986 TS$_6$ family members had
featureless spectra and high slopes resulting in a classification for
most as D-types, while the few with lower slopes were placed in the C-
and P-classes. Eurybates members, however, have markedly different
spectra with low to moderate slopes splitting the classifications
evenly between the C- or P-classes ({\it Fornasier et al.} 2007, {\it DeLuise et
al.} 2010).  {\it Roig et al.} (2008b) used the SDSS data to investigate
asteroid families in the Jupiter Trojan population, and found that the
families in the Trojan populations account for the differences in the
compositional distributions between the L4 and L5 clouds.  In particular,
the families in the L4 cloud show an abundance of C- and P-type
objects not reflected in the L5 families or the background populations
in either cloud.

The TNO population covers a much larger volume of space than any of
the populations of objects closer to the Sun, but is also estimated to
contain over 1000 times the mass of the Main Belt.  Collisions
resulting in catastrophic disruptions are believed to have occurred at
least twice in the TNO population.  Pluto's five (or potentially more;
{\it Weaver et al.} 2006) satellites speak to a massive collision, which
will be a key area of investigation of the New Horizons flyby of the
Pluto system.

The dwarf planet Haumea is highly elongated with a very short
rotational period ($\sim 4$ hours), is orbited by two small
satellites, has a relatively high density, and has a spectrum that is
consistent with nearly pure water ice. These properties are thought to
be the result of a mantle-shattering collisional event, though the
details of this event remain contentious.  A group of TNOs with colors
substantially bluer than the typical neutral-to-ultra-red surfaces of
the Kuiper Belt, all sharing high inclinations similar to Haumea, has
been identified as a collisional family produced by this event ({\it
  Brown et al.} 2007). Because of the orbital velocity regimes in
trans-Neptunian space, collisional families are in general unlikely to
be identified there through dynamics alone (as they are in the
asteroid belt). It was only through the unique composition of the
family members (akin to the extremely distinct photometric properties
of the Vesta family members) that the Haumea family could be readily
identified.  This implies that more collisional families may be hiding
in the TNO population, but cannot be spotted by orbits alone.

Although massive collisions dominated the solar system environment
during the epoch of planet building, they also played an important
role in shaping all of the major populations of small bodies during
the subsequent $\sim 4$ billion years.  Collisional evolution,
although stochastic in nature, was a major determinant in the
structure of the solar system we see today (see the chapter by {\it
  Bottke et al.} in this volume).

\bigskip
\section{\textbf{OPEN PROBLEMS AND FUTURE PROSPECTS}}
\bigskip

The effect of space weathering on asteroidal surfaces still remains an
important topic for future exploration (see the chapter by {\it Brunetto et
al.} in this volume for further discussion).  The range of studies
carried out so far have found a wide dispersion of results, both in
terms of the timescale of weathering and the specific effects on
various taxonomic classes.  Studies to date have relied on the
assumption that all asteroids with similar taxonomic types have
identical mineralogical compositions, or have been based on a single
pair of families known to be compositionally identical but with
different ages (i.e. Karin and Koronis).  As deeper surveys and
dynamical analysis techniques begin to increase the number of
identified cases of family-within-a-family, physical studies of these
interesting populations will allow better measurements of the specific
effects and timing of space weathering processes.

Advances in physical measurements of asteroid families have not
uniformly addressed the various parameters needed for a robust
scientific investigation.  In particular, there has been only a
nominal advancement in the measurement of asteroid masses and
densities, owing to the difficulty in determining these parameters.
{\it Carry} (2012) compiled known measurements of asteroid densities into a
comprehensive list, however only a handful of families have more than
one member represented, and for most the only measurement is of the
largest remnant body.  A larger survey of densities including many
family members over a range of sizes would enable testing of family
formation conditions via reaccretion (cf. chapter by {\it Michel et al.} in
this volume), as well as improve family ages derived from numerical
simulations of gravitational and Yarkovsky orbital evolution
(cf. chapter by {\it Nesvorn\'{y} et al.} in this volume).  This could be
accomplished by a more comprehensive search for binaries, or through
modeling of gravitational perturbations to asteroid orbits detectable
in next-generation astrometric catalogs.

One highly-anticipated survey will be carried out by the European
Space Agency's Gaia mission ({\it Hestroffer et al.}, 2010), which is
expected to provide spectral characterization of all objects down to
an apparent magnitude of $V=20$, including many asteroid family
members ({\it Campins et al.} 2012; {\it Delbo et al.} 2012; {\it
  Cellino \& Dell'Oro}, 2012).  This dramatic increase in the number
of characterized family members will enable a host of new
investigations of the composition and differentiation of family parent
bodies.  By spectrally probing bodies as small as $\sim2~$km,
inhomogeneities in asteroid family composition may begin to be
revealed.  Another key scientific product will be refined astrometry
for all asteroids which can be searched for gravitational
perturbations and used to determine the masses of the largest
asteroids ({\it Mignard et al.}, 2007).  This wealth of new data will
feed into theoretical models and numerical simulations, allowing us to
improve the ages determined from orbital evolution simulations
(e.g. {\it Masiero et al.}, 2012; {\it Carruba et al.}, 2014).

The collision events that form families are very rare, and the
probability of a massive collision in the next ten, hundred, or
thousand years is vanishingly small (see chapter by {\it Bottke et al.} in
this volume).  However, a new class of active Main Belt objects has
recently been identified that may none-the-less provide a window into
the collisional environment of the solar system (see the chapter by
{\it Jewitt et al.} in this volume for further discussion).  While some of
these objects show repeated activity indicative of a cometary nature
complete with sub-surface volatiles, others are best explained by
impact events.  In particular, the observed outburst events of P/2010
A2 ({\it Jewitt et al.} 2010) and (596) Scheila ({\it Jewitt et al.} 2011) are
consistent with impacts by very small ($D<50~$m) asteroids.  As
current and future sky surveys probe to smaller diameters in the Main
Belt, the frequency these events will be observed, or even predicted
in advance, will increase.  Study of these small scale disruptions
offers an important constraint on impact physics in the low-gravity
environment of asteroids, and provides test cases for comparing to the
small-scale impact experiments that can be performed in Earth-based
laboratories.

Recently, {\it Reddy et al.} (2014) presented evidence that shock darkening
may play a role in altering the spectroscopic properties of chondritic
materials in the Baptistina and Flora families.  If evidence for this
effect is seen in other families across a range of compositions, this
technique may provide a method for determining the conditions of
family-forming impacts in the Main Belt, and thus provide better
constraints on the ages of families.  This may also help explain some
of the differences observed in space weathering studies that find
albedo changes happen very quickly while spectral changes have long
timescales.

As new data sets have rapidly increased our ability to characterize
asteroid family members, and through this the parent body from which
they originated, one glaring question remains at the forefront of the
field: where are the families of differentiated parent bodies that
were completely disrupted?  Vesta and its family members have given us
a excellent example of the composition and resultant albedo, color,
and spectral properties of the crust of a differentiated body (see the
chapter by {\it Russell et al.} in this volume for further discussion).
However, searches for objects with similar properties in different
regions of the Main Belt have yielded no significant populations of
these basaltic crust and mantle fragments that should be leftover from
these collisions.  On the other hand, data from iron meteorites has
indicated that differentiation of protoplanets, if not common, at a
minimum happened multiple times (see chapters by {\it Elkins-Tanton et
al.} and {\it Scott et al.} in this volume for further discussion).  While it
is possible that these collisions happened at the earliest stages of
the solar system's formation and the evidence (in the form of
families) was erased during the epoch of giant planet migration that
is the foundation of the 'Nice' model (see chapter by {\it Morbidelli et
al.} in this volume), a dynamical explanation of this problem would
need to preserve in the Main Belt the core material that still falls
to Earth today.  Conversely, if these impacts happened after the last
great shake up of the solar system, a mineralogical explanation for
how metallic cores could form without leaving a basaltic 'residue' in
the Main Belt is required.  The limitations on what {\em could not}
have happened that are being set by current surveys are just as
important as our discoveries of what did happen.

The next decade of large surveys, both ground- and space-based,
promises to expand our knowledge of asteroid physical properties by
potentially another order of magnitude beyond what is known today.  In
this data-rich environment, family research will focus not just on
individual families and their place in the Main Belt, but on specific
sub-groupings within families: on the knots, clumps, and collisional
cascade fragments that trace the disruption and evolutionary dynamics
that families have undergone.  As the sizes of objects probed reach
smaller and smaller, we can expect to find more young families like
Karin and Iannini that can be directly backward-integrated to a
specific time of collision, improving our statistics of collisions in
the last 10 million years.  As catalogs increase in size, we can also
expect to more frequently have characterization data of objects both
before and after they undergo catastrophic disruption.  This will
enable us to test our impact physics models on scales not achievable
on Earth.  Additionally, we will begin to see a time when we routinely
use remote sensing data of asteroids to not just associate families
but assess the mineralogy of family members as a probe of the parent
body.  Asteroid family physical properties, numerical simulations, and
evolutionary theory will leapfrog off of each other, pushing forward
our understanding of the asteroids and of the whole solar system.

\textbf{ Acknowledgments.} JRM was partially supported by a grant from
the NASA Planetary Geology and Geophysics Program. FED was supported
by NASA through the Hubble Fellowship grant HST-HF-51319.01-A, awarded
by the Space Telescope Science Institute, which is operated by the
Association of Universities for Research in Astronomy, Inc., for NASA,
under contract NAS 5-26555.  This publication makes use of data
products from the Wide-field Infrared Survey Explorer, which is a
joint project of the University of California, Los Angeles, and the
Jet Propulsion Laboratory/California Institute of Technology, funded
by the National Aeronautics and Space Administration. This publication
also makes use of data products from NEOWISE, which is a project of
the Jet Propulsion Laboratory/California Institute of Technology,
funded by the Planetary Science Division of the National Aeronautics
and Space Administration.  Funding for the creation and distribution
of the SDSS Archive has been provided by the Alfred P. Sloan
Foundation, the Participating Institutions, the National Aeronautics
and Space Administration, the National Science Foundation, the
U.S. Department of Energy, the Japanese Monbukagakusho, and the Max
Planck Society. The SDSS Web site is http://www.sdss.org/.  The SDSS
is managed by the Astrophysical Research Consortium (ARC) for the
Participating Institutions. The Participating Institutions are The
University of Chicago, Fermilab, the Institute for Advanced Study, the
Japan Participation Group, The Johns Hopkins University, the Korean
Scientist Group, Los Alamos National Laboratory, the
Max-Planck-Institute for Astronomy (MPIA), the Max-Planck-Institute
for Astrophysics (MPA), New Mexico State University, University of
Pittsburgh, University of Portsmouth, Princeton University, the United
States Naval Observatory, and the University of Washington.  This
research is based on observations with AKARI, a JAXA project with the
participation of ESA. The authors wish to thank Bobby Bus, Beth Ellen
Clark, Heather Kaluna, and Pierre Vernazza for providing proprietary
spectra to include in Figure 1.  We also thank the referees, Alberto
Cellino and Bobby Bus, and editor Patrick Michel for helpful comments
that improved this chapter.

\begin{deluxetable}{rcrcrcrccrcccr}
\tabletypesize{\tiny}
\tablecaption{Physical properties of asteroid families \label{tbl.fams}}
\tablewidth{0pt}

\tablehead{Number & Name & PDS ID & $<p_V>$ & $n_{p_V}$ & $<p_{NIR}>$ & $n_{p_{NIR}}$ & $<a^{\star}>$ & $<i-z>$ & $n_{SDSS}$ & SFD slope & SFD range (km)&Tax&$n_{tax}$\\}

\startdata
     2 &           Pallas & 801  &0.134+/-0.026 &    49 & 0.114+/-0.047 &    10 & -0.14+/-0.03 & -0.01+/-0.08 &     13 &   ...+/-...   &     ...     &   B &   8 \\
     3 &             Juno & 501  &0.262+/-0.054 &   125 & 0.488+/-0.000 &     8 &  0.08+/-0.05 & -0.03+/-0.07 &     87 &-2.427+/-0.108 &   2.7-7.3   &  Sq &   1 \\
     4 &            Vesta & 401  &0.363+/-0.088 &  1900 & 0.465+/-0.156 &    54 &  0.12+/-0.05 & -0.26+/-0.10 &   2148 &-3.417+/-0.030 &   2.5-11.9  &   V &  49 \\
     8 &            Flora & 402  &0.305+/-0.064 &  1330 & 0.440+/-0.082 &   142 &  0.13+/-0.06 & -0.04+/-0.07 &    922 &-2.692+/-0.030 &   2.8-12.2  &   S &  74 \\
    10 &           Hygiea & 601  &0.070+/-0.018 &  1951 & 0.065+/-0.028 &     3 & -0.11+/-0.05 &  0.01+/-0.07 &    606 &-3.883+/-0.040 &   6.1-19.3  &   C &   1 \\
    15 &          Eunomia & 502  &0.270+/-0.059 &  1448 & 0.374+/-0.088 &   148 &  0.13+/-0.05 & -0.03+/-0.06 &    798 &-3.091+/-0.033 &   4.4-17.6  &   S &  30 \\
    20 &         Massalia & 404  &0.247+/-0.053 &   214 &   ...+/-...   &     0 &  0.07+/-0.05 & -0.04+/-0.08 &    386 &-3.544+/-0.140 &   2.0-4.1   &   S &   1 \\
    24 &           Themis & 602  &0.068+/-0.017 &  2218 & 0.074+/-0.030 &    86 & -0.11+/-0.04 &  0.01+/-0.06 &    640 &-2.313+/-0.017 &   7.3-55.6  &   C &   7 \\
    25 &          Phocaea & 701  &0.290+/-0.066 &   715 & 0.355+/-0.203 &   119 &  0.10+/-0.11 & -0.04+/-0.08 &    252 &-2.663+/-0.039 &   3.5-14.6  &   S &  39 \\
    27 &          Euterpe & 410  &0.270+/-0.062 &    45 & 0.493+/-0.000 &     1 &  0.11+/-0.05 & -0.04+/-0.08 &     42 &   ...+/-...   &     ...     &   S &   1 \\
    31 &       Euphrosyne & 901  &0.059+/-0.013 &   742 & 0.082+/-0.173 &     5 & -0.08+/-0.05 &  0.04+/-0.06 &    169 &-4.687+/-0.082 &   7.3-18.4  &  Cb &   1 \\
    44 &     Nysa(Polana) & 405  &0.289+/-0.074 &  1345 & 0.356+/-0.059 &    22 &  0.13+/-0.06 & -0.03+/-0.07 &   1544 &-3.083+/-0.030 &   2.5-13.0  & ... &   0 \\
    46 &           Hestia & 503  &0.267+/-0.049 &    28 & 0.068+/-0.000 &     1 &  0.14+/-0.04 & -0.03+/-0.04 &     22 &   ...+/-...   &     ...     &  Xc &   1 \\
    81 &      Terpsichore & 622  &0.050+/-0.010 &    57 & 0.053+/-0.000 &     1 & -0.08+/-0.03 &  0.08+/-0.10 &     15 &-4.371+/-0.508 &   5.0-7.3   &  Cb &   1 \\
    84 &             Klio & 413  &0.059+/-0.014 &   107 & 0.089+/-0.031 &     3 & -0.06+/-0.04 &  0.05+/-0.08 &     23 &-2.478+/-0.129 &   3.3-8.0   &  Ch &   1 \\
    87 &           Sylvia & 603  &0.051+/-0.012 &   121 & 0.082+/-0.000 &     1 & -0.07+/-0.05 &  0.09+/-0.11 &     20 &-3.339+/-0.232 &   7.0-12.4  &   X &   1 \\
    89 &            Julia & 540  &0.225+/-0.036 &     2 & 0.339+/-0.000 &     2 &  0.05+/-0.02 &  0.12+/-0.07 &      3 &   ...+/-...   &     ...     &  Ld &   1 \\
    96 &            Aegle & 630  &0.072+/-0.013 &    83 & 0.102+/-0.000 &     1 &  0.02+/-0.05 &  0.10+/-0.08 &     23 &-3.252+/-0.305 &   6.8-11.4  &   T &   1 \\
   128 &          Nemesis & 504  &0.072+/-0.019 &   347 & 0.071+/-0.000 &     1 & -0.08+/-0.06 &  0.03+/-0.07 &    109 &-4.320+/-0.121 &   3.8-8.4   &   C &   2 \\
   137 &         Meliboea & 604  &0.060+/-0.013 &   163 & 0.050+/-0.029 &    12 & -0.10+/-0.08 &  0.05+/-0.07 &     53 &-1.513+/-0.051 &   8.0-39.4  &  Ch &  11 \\
   142 &     Polana(Nysa) & n/a  &0.056+/-0.012 &  1130 & 0.061+/-0.006 &     3 & -0.12+/-0.10 &  0.00+/-0.08 &    375 &-3.177+/-0.041 &   3.6-12.4  &   B &   3 \\
   144 &          Vibilia & 529  &0.065+/-0.011 &   180 &   ...+/-...   &     0 & -0.10+/-0.04 &  0.04+/-0.07 &     35 &-3.137+/-0.135 &   4.5-9.7   &  Ch &   1 \\
   145 &           Adeona & 505  &0.060+/-0.011 &   874 & 0.068+/-0.078 &    19 & -0.11+/-0.08 &  0.04+/-0.07 &    274 &-2.854+/-0.037 &   5.3-23.2  &  Ch &  12 \\
   148 &           Gallia & 802  &0.251+/-0.059 &    24 &   ...+/-...   &     0 &  0.10+/-0.03 & -0.06+/-0.06 &     15 &   ...+/-...   &     ...     &  Sl &   1 \\
   158 &          Koronis & 605  &0.238+/-0.051 &  1089 & 0.325+/-0.059 &    67 &  0.09+/-0.07 & -0.02+/-0.08 &    810 &-2.451+/-0.026 &   5.1-32.0  &   S &  34 \\
   163 &          Erigone & 406  &0.051+/-0.010 &   716 & 0.061+/-0.013 &    17 & -0.08+/-0.10 &  0.05+/-0.08 &    201 &-3.215+/-0.050 &   3.3-11.9  &  Ch &   1 \\
   170 &            Maria & 506  &0.255+/-0.061 &  1361 & 0.370+/-0.068 &    69 &  0.12+/-0.05 & -0.02+/-0.07 &    809 &-2.637+/-0.025 &   3.5-24.8  &   S &  23 \\
   173 &              Ino & 522  &0.244+/-0.069 &    90 &   ...+/-...   &     0 &  0.09+/-0.05 & -0.06+/-0.07 &     93 &-3.141+/-0.268 &   2.7-4.6   & ... &   0 \\
   221 &              Eos & 606  &0.163+/-0.035 &  3509 & 0.180+/-0.053 &   205 &  0.05+/-0.05 &  0.03+/-0.07 &   1692 &-2.222+/-0.013 &   5.6-47.3  &   K &  30 \\
   283 &             Emma & 607  &0.047+/-0.011 &   260 & 0.118+/-0.113 &     2 & -0.07+/-0.06 &  0.04+/-0.08 &     75 &-3.442+/-0.113 &   6.7-15.3  &   C &   1 \\
   293 &         Brasilia & 608  &0.174+/-0.048 &   110 & 0.224+/-0.017 &     4 & -0.04+/-0.04 &  0.05+/-0.07 &     78 &-3.462+/-0.243 &   3.7-6.4   &   X &   2 \\
   298 &       Baptistina & 403  &0.179+/-0.056 &   581 & 0.390+/-0.198 &    25 &  0.01+/-0.09 & -0.02+/-0.09 &    321 &-3.254+/-0.063 &   2.5-7.2   &   S &   9 \\
   302 &         Clarissa & 407  &0.048+/-0.010 &    44 &   ...+/-...   &     0 & -0.14+/-0.04 &  0.00+/-0.06 &     10 &   ...+/-...   &     ...     &   F &   1 \\
   313 &         Chaldaea & 415  &0.063+/-0.017 &   169 & 0.083+/-0.039 &    11 & -0.10+/-0.05 &  0.06+/-0.07 &     34 &-3.058+/-0.119 &   3.7-8.6   &   C &   3 \\
   322 &            Phaeo & 530  &0.059+/-0.015 &    99 & 0.181+/-0.000 &     2 &  0.00+/-0.06 &  0.05+/-0.08 &     34 &-2.852+/-0.195 &   5.1-10.1  &   D &   1 \\
   329 &             Svea & 416  &0.050+/-0.009 &    30 & 0.134+/-0.166 &     2 & -0.04+/-0.06 &  0.04+/-0.05 &      8 &   ...+/-...   &     ...     &   C &   1 \\
   363 &            Padua & 507  &0.069+/-0.015 &   427 & 0.067+/-0.022 &     5 & -0.04+/-0.06 &  0.05+/-0.07 &    135 &-2.588+/-0.053 &   4.5-16.4  &   X &  10 \\
   369 &            Aeria & 539  &0.180+/-0.011 &    22 & 0.266+/-0.000 &     6 & -0.05+/-0.04 &  0.03+/-0.09 &     18 &   ...+/-...   &     ...     &   M &   1 \\
   375 &           Ursula & 631  &0.061+/-0.014 &   600 & 0.083+/-0.047 &    15 & -0.05+/-0.08 &  0.06+/-0.09 &    194 &-2.677+/-0.046 &   7.3-27.2  & ... &   0 \\
   396 &           Aeolia & 508  &0.107+/-0.022 &    43 & 0.115+/-0.000 &     1 & -0.04+/-0.05 &  0.04+/-0.03 &     10 &   ...+/-...   &     ...     &  Xe &   1 \\
   410 &          Chloris & 509  &0.084+/-0.031 &   171 & 0.081+/-0.095 &    10 & -0.05+/-0.08 &  0.06+/-0.05 &     59 &-2.814+/-0.103 &   5.2-14.0  &   C &   8 \\
   434 &         Hungaria & 003  &0.456+/-0.217 &   527 & 0.727+/-1.692 &     9 & -0.01+/-0.08 &  0.04+/-0.10 &    636 &   ...+/-...   &     ...     &  Xe &  14 \\
   480 &            Hansa & 803  &0.269+/-0.067 &   316 & 0.377+/-0.068 &     9 &  0.10+/-0.05 & -0.06+/-0.07 &    134 &-3.378+/-0.104 &   3.4-7.9   &   S &   2 \\
   490 &          Veritas & 609  &0.066+/-0.016 &   697 & 0.068+/-0.011 &     3 & -0.07+/-0.04 &  0.05+/-0.08 &    207 &-2.744+/-0.043 &   5.8-22.6  &  Ch &   8 \\
   495 &  Eulalia(Polana) & n/a  &0.057+/-0.012 &  2008 & 0.066+/-0.029 &    15 & -0.12+/-0.05 &  0.01+/-0.07 &    531 &-2.687+/-0.021 &   3.0-18.4  &   C &  13 \\
   569 &             Misa & 510  &0.052+/-0.013 &   287 & 0.064+/-0.027 &     3 & -0.07+/-0.07 &  0.03+/-0.08 &     79 &-2.508+/-0.067 &   4.0-13.3  &  Ch &   1 \\
   589 &          Croatia & 638  &0.054+/-0.010 &    99 & 0.068+/-0.000 &     1 & -0.05+/-0.04 &  0.02+/-0.09 &     22 &-3.383+/-0.224 &   5.8-11.1  &   X &   1 \\
   606 &         Brangane & 511  &0.112+/-0.028 &    44 & 0.137+/-0.000 &     2 &  0.07+/-0.04 &  0.09+/-0.06 &     20 &   ...+/-...   &     ...     &   L &   1 \\
   618 &         Elfriede & 632  &0.052+/-0.012 &    36 & 0.063+/-0.000 &     1 & -0.09+/-0.05 &  0.06+/-0.05 &      8 &   ...+/-...   &     ...     &   C &   1 \\
   623 &         Chimaera & 414  &0.054+/-0.012 &    63 & 0.049+/-0.000 &     3 & -0.05+/-0.03 &  0.09+/-0.10 &     20 &-2.586+/-0.235 &   3.8-7.2   &  XC &   1 \\
   627 &           Charis & 616  &0.071+/-0.010 &    39 & 0.264+/-0.000 &     2 &  0.16+/-0.06 &  0.04+/-0.08 &     59 &   ...+/-...   &     ...     &   X &   1 \\
   656 &           Beagle & 620  &0.080+/-0.014 &    30 & 0.070+/-0.006 &     5 & -0.12+/-0.03 &  0.01+/-0.06 &     16 &   ...+/-...   &     ...     &   X &   1 \\
   668 &             Dora & 512  &0.056+/-0.012 &   667 & 0.047+/-0.017 &    17 & -0.12+/-0.07 &  0.04+/-0.07 &    175 &-2.610+/-0.043 &   5.7-21.6  &  Ch &  29 \\
   686 &         Gersuind & 804  &0.145+/-0.037 &   106 & 0.328+/-0.000 &     1 &  0.09+/-0.07 &  0.08+/-0.07 &     36 &-2.653+/-0.158 &   3.8-8.0   & ... &   0 \\
   702 &           Alauda & 902  &0.066+/-0.015 &   687 & 0.071+/-0.013 &    26 & -0.10+/-0.06 &  0.03+/-0.07 &    197 &-2.707+/-0.042 &   9.2-36.7  & ... &   0 \\
   709 &        Fringilla & 623  &0.050+/-0.014 &    51 & 0.212+/-0.788 &     2 & -0.03+/-0.06 &  0.09+/-0.08 &     24 &   ...+/-...   &     ...     &   X &   1 \\
   729 &         Watsonia & 537  &0.134+/-0.019 &    51 & 0.181+/-0.000 &     2 &  0.08+/-0.04 &  0.07+/-0.02 &     12 &   ...+/-...   &     ...     &   L &   2 \\
   752 &        Sulamitis & 408  &0.055+/-0.011 &   134 & 0.048+/-0.000 &     1 & -0.07+/-0.07 &  0.06+/-0.11 &     33 &-2.417+/-0.126 &   3.8-8.8   &  Ch &   1 \\
   778 &        Theobalda & 617  &0.062+/-0.016 &   107 & 0.070+/-0.000 &     1 & -0.16+/-0.03 &  0.00+/-0.05 &     28 &-3.097+/-0.185 &   6.5-13.3  & ... &   0 \\
   780 &          Armenia & 905  &0.056+/-0.013 &    28 & 0.060+/-0.000 &     2 & -0.05+/-0.01 &  0.03+/-0.04 &      4 &   ...+/-...   &     ...     &   C &   1 \\
   808 &           Merxia & 513  &0.234+/-0.054 &    93 & 0.347+/-0.030 &     3 &  0.08+/-0.05 & -0.08+/-0.08 &     98 &-2.662+/-0.190 &   3.2-6.7   &   S &   6 \\
   832 &            Karin & 610  &0.178+/-0.031 &    18 & 0.294+/-0.000 &     1 &  0.03+/-0.05 & -0.01+/-0.07 &     33 &   ...+/-...   &     ...     &   S &  47 \\
   845 &            Naema & 611  &0.064+/-0.012 &   155 & 0.055+/-0.000 &     1 & -0.10+/-0.11 &  0.04+/-0.06 &     45 &-4.274+/-0.220 &   6.0-10.6  &   C &   1 \\
   847 &            Agnia & 514  &0.238+/-0.060 &   110 & 0.389+/-0.015 &     3 &  0.04+/-0.05 & -0.07+/-0.08 &    179 &-3.055+/-0.165 &   3.8-8.0   &   S &   8 \\
   909 &             Ulla & 903  &0.048+/-0.009 &    19 &   ...+/-...   &     0 & -0.07+/-0.02 &  0.02+/-0.06 &      3 &   ...+/-...   &     ...     &   X &   1 \\
   918 &             Itha & 633  &0.239+/-0.056 &    28 & 0.353+/-0.084 &     8 &  0.14+/-0.04 & -0.01+/-0.05 &     14 &   ...+/-...   &     ...     &  Sl &   4 \\
   945 &        Barcelona & 805  &0.290+/-0.064 &    52 & 0.510+/-0.000 &     1 &  0.09+/-0.05 & -0.11+/-0.09 &     21 &   ...+/-...   &     ...     &  Sq &   1 \\
  1128 &           Astrid & 515  &0.045+/-0.010 &   213 & 0.046+/-0.000 &     1 & -0.07+/-0.05 &  0.08+/-0.08 &     33 &-2.567+/-0.080 &   3.5-11.1  &   C &   5 \\
  1189 &         Terentia & 618  &0.064+/-0.012 &    13 & 0.058+/-0.000 &     1 & -0.01+/-0.00 &  0.09+/-0.00 &      1 &   ...+/-...   &     ...     &  Ch &   1 \\
  1222 &             Tina & 806  &0.128+/-0.042 &    26 & 0.137+/-0.000 &     1 &  0.01+/-0.00 &  0.03+/-0.05 &      3 &   ...+/-...   &     ...     &   X &   1 \\
  1270 &           Datura & 411  &0.288+/-0.000 &     1 &   ...+/-...   &     0 &  0.01+/-0.00 & -0.10+/-0.00 &      1 &   ...+/-...   &     ...     &   S &   4 \\
  1272 &           Gefion & 516  &0.267+/-0.058 &   737 & 0.350+/-0.154 &    25 &  0.10+/-0.06 & -0.02+/-0.07 &    523 &-3.262+/-0.050 &   4.0-14.3  &   S &  32 \\
  1303 &          Luthera & 904  &0.050+/-0.009 &   125 & 0.078+/-0.000 &     1 &  0.01+/-0.04 &  0.10+/-0.07 &     39 &-3.955+/-0.264 &   7.9-13.0  & ... &   0 \\
  1332 &         Marconia & 636  &0.042+/-0.008 &     6 & 0.080+/-0.000 &     1 &   ...+/-...  &   ...+/-...  &      0 &   ...+/-...   &     ...     &   L &   1 \\
  1400 &           Tirela & 612  &0.239+/-0.057 &   419 & 0.073+/-0.095 &     7 &  0.15+/-0.09 &  0.08+/-0.08 &    225 &-3.454+/-0.073 &   4.8-14.3  &  Ld &  10 \\
  1484 &         Postrema & 541  &0.047+/-0.010 &    30 & 0.051+/-0.006 &     3 & -0.05+/-0.05 &  0.02+/-0.06 &      5 &   ...+/-...   &     ...     &   B &   1 \\
  1658 &            Innes & 518  &0.259+/-0.057 &   195 & 0.318+/-0.149 &     8 &  0.09+/-0.06 & -0.04+/-0.07 &    163 &-3.385+/-0.128 &   3.1-7.0   &   S &   2 \\
  1668 &            Hanna & 533  &0.052+/-0.011 &   102 & 0.048+/-0.000 &     1 & -0.09+/-0.03 &  0.08+/-0.05 &     20 &-3.502+/-0.240 &   3.9-7.2   & ... &   0 \\
  1726 &      Hoffmeister & 519  &0.047+/-0.010 &   609 & 0.052+/-0.013 &     4 & -0.10+/-0.05 &  0.03+/-0.08 &    145 &-2.856+/-0.047 &   4.5-17.2  &   C &   9 \\
  1892 &         Lucienne & 409  &0.228+/-0.029 &    19 &   ...+/-...   &     0 &  0.09+/-0.03 &  0.00+/-0.09 &     10 &   ...+/-...   &     ...     &   S &   1 \\
  2085 &            Henan & 542  &0.227+/-0.065 &   106 & 0.165+/-0.000 &     3 &  0.11+/-0.06 & -0.02+/-0.09 &    163 &-3.667+/-0.276 &   3.4-5.8   &   L &   1 \\
  2262 &         Mitidika & 531  &0.066+/-0.014 &   279 & 0.060+/-0.013 &    10 & -0.11+/-0.06 &  0.01+/-0.08 &     71 &-2.832+/-0.091 &   4.6-11.4  & ... &   0 \\
  2344 &           Xizang & 536  &0.134+/-0.044 &    18 &   ...+/-...   &     0 &  0.04+/-0.05 &  0.08+/-0.07 &     12 &   ...+/-...   &     ...     & ... &   0 \\
  2384 &         Schulhof & 525  &0.280+/-0.000 &     1 & 0.289+/-0.000 &     4 &  0.08+/-0.07 & -0.05+/-0.03 &      2 &   ...+/-...   &     ...     & ... &   0 \\
  2732 &             Witt & 535  &0.261+/-0.053 &   124 & 0.347+/-0.018 &     3 &  0.15+/-0.07 &  0.01+/-0.08 &    190 &-3.114+/-0.173 &   3.8-7.7   & ... &   0 \\
  2782 &         Leonidas & 528  &0.068+/-0.013 &    67 &   ...+/-...   &     0 & -0.10+/-0.07 &  0.04+/-0.07 &     22 &-2.142+/-0.245 &   5.1-9.4   & ... &   0 \\
  3152 &            Jones & 538  &0.054+/-0.004 &    14 & 0.061+/-0.000 &     1 & -0.02+/-0.00 &  0.04+/-0.01 &      2 &   ...+/-...   &     ...     &   T &   1 \\
  3438 &        Inarrados & 634  &0.067+/-0.015 &    24 & 0.069+/-0.000 &     1 & -0.10+/-0.03 &  0.05+/-0.05 &     11 &   ...+/-...   &     ...     & ... &   0 \\
  3556 &        Lixiaohua & 613  &0.044+/-0.009 &   367 & 0.053+/-0.011 &     3 & -0.06+/-0.05 &  0.04+/-0.09 &     95 &-3.499+/-0.098 &   6.8-15.7  & ... &   0 \\
  3811 &            Karma & 534  &0.054+/-0.010 &    78 &   ...+/-...   &     0 & -0.06+/-0.08 &  0.03+/-0.07 &     20 &-2.659+/-0.241 &   3.7-6.7   & ... &   0 \\
  3815 &            Konig & 517  &0.049+/-0.011 &   135 &   ...+/-...   &     0 & -0.08+/-0.02 &  0.06+/-0.06 &     25 &-3.032+/-0.150 &   4.1-9.2   & ... &   0 \\
  4203 &          Brucato & 807  &0.064+/-0.017 &   233 & 0.041+/-0.009 &     2 & -0.11+/-0.05 &  0.05+/-0.07 &     45 &-3.604+/-0.109 &   5.3-13.3  & ... &   0 \\
  4652 &          Iannini & 520  &0.309+/-0.033 &    18 &   ...+/-...   &     0 &  0.01+/-0.04 & -0.08+/-0.06 &     21 &   ...+/-...   &     ...     & ... &   0 \\
  5438 &            Lorre & 527  &0.054+/-0.002 &     2 &   ...+/-...   &     0 &   ...+/-...  &   ...+/-...  &      0 &   ...+/-...   &     ...     &   C &   1 \\
  5567 &          Durisen & 624  &0.043+/-0.010 &    16 & 0.050+/-0.000 &     1 & -0.02+/-0.03 &  0.11+/-0.05 &      4 &   ...+/-...   &     ...     & ... &   0 \\
  5614 &         Yakovlev & 625  &0.046+/-0.008 &    25 &   ...+/-...   &     0 & -0.08+/-0.00 &  0.07+/-0.01 &      2 &   ...+/-...   &     ...     & ... &   0 \\
  7353 &           Kazuia & 532  &0.214+/-0.025 &    13 &   ...+/-...   &     0 &  0.06+/-0.08 & -0.03+/-0.02 &      3 &   ...+/-...   &     ...     &   L &   1 \\
  7468 &          Anfimov & 635  &0.166+/-0.061 &    10 &   ...+/-...   &     0 &  0.23+/-0.03 & -0.14+/-0.09 &     13 &   ...+/-...   &     ...     & ... &   0 \\
  7481 &      SanMarcello & 626  &0.194+/-0.054 &    27 &   ...+/-...   &     0 & -0.05+/-0.04 &  0.04+/-0.06 &     18 &   ...+/-...   &     ...     & ... &   0 \\
  9506 &        Telramund & 614  &0.237+/-0.063 &    46 &   ...+/-...   &     0 &  0.09+/-0.05 & -0.02+/-0.09 &     33 &   ...+/-...   &     ...     & ... &   0 \\
 10811 &              Lau & 619  &0.274+/-0.000 &     1 &   ...+/-...   &     0 &  0.26+/-0.00 & -0.18+/-0.00 &      1 &   ...+/-...   &     ...     & ... &   0 \\
 14627 &     Emilkowalski & 523  &0.149+/-0.046 &     4 &   ...+/-...   &     0 &  0.06+/-0.00 &  0.11+/-0.00 &      1 &   ...+/-...   &     ...     & ... &   0 \\
 15454 &          1998YB3 & 627  &0.054+/-0.006 &    11 &   ...+/-...   &     0 & -0.11+/-0.03 &  0.04+/-0.03 &      7 &   ...+/-...   &     ...     & ... &   0 \\
 15477 &          1999CG1 & 628  &0.095+/-0.030 &    68 &   ...+/-...   &     0 &  0.04+/-0.04 &  0.05+/-0.06 &     25 &-4.768+/-0.602 &   4.3-6.0   & ... &   0 \\
 16598 &          1992YC2 & 524  &  ...+/-...   &     0 &   ...+/-...   &     0 &  0.10+/-0.00 & -0.09+/-0.00 &      1 &   ...+/-...   &     ...     &  Sq &   1 \\
 18405 &         1993FY12 & 615  &0.173+/-0.049 &    15 &   ...+/-...   &     0 & -0.04+/-0.03 &  0.07+/-0.08 &     12 &   ...+/-...   &     ...     & ... &   0 \\
 36256 &         1999XT17 & 629  &0.203+/-0.066 &    13 &   ...+/-...   &     0 &  0.21+/-0.07 & -0.04+/-0.05 &      6 &   ...+/-...   &     ...     & ... &   0 \\
 53546 &          2000BY6 & 526  &0.139+/-0.000 &     1 &   ...+/-...   &     0 &  0.04+/-0.03 &  0.06+/-0.04 &      9 &   ...+/-...   &     ...     & ... &   0 \\
106302 &         2000UJ87 & 637  &0.045+/-0.004 &    12 &   ...+/-...   &     0 & -0.10+/-0.04 &  0.05+/-0.04 &      3 &   ...+/-...   &     ...     & ... &   0 \\
\enddata

\end{deluxetable}

\clearpage

\bigskip

\centerline\textbf{ REFERENCES}
\bigskip
\parskip=0pt
{\small
\baselineskip=11pt

\refs Al\'{i}-Lagoa, V., de Le\'{o}n, J., Licandro, et al. (2013) Physical properties of B-type asteroids from WISE data. {\em Astron. Astrophys., 554}, A71.

\refs Assandri, M. C. \& Gil-Hutton, R. (2008) Surface composition of Hungaria asteroids from the analysis of the Sloan Digital Sky Survey. {\em Astron. \& Astrophys., 488}, 339.

\refs Beaug\'{e}, C. \& Roig, F. (2001) A Semianalytical Model for the Motion of the Trojan Asteroids: Proper Elements and Families. {\em Icarus, 153}, 391.

\refs Bell, J. F. (1988) A probable asteroidal parent body for the CO or CV chondrites. {\em Meteoritics, 23}, 256.


\refs Binzel, R. P. \& Xu, S. (1993) Chips off of asteroid 4 Vesta - Evidence for the parent body of basaltic achondrite meteorites. {\em Science, 260}, 186.


\refs Bottke, W. F., Vokrouhlick\'{y}, D., Minton, D., et al. (2012) An Archaean heavy bombardment from a destabilized extension of the asteroid belt. {\em Nature, 485}, 78.

\refs Bottke, W. F., Vokrouhlick\'{y}, D., Nesvorn\'{y}, D., (2007) An asteroid breakup 160Myr ago as the probable source of the K/T impactor. {\em Nature, 449}, 48.

\refs Bottke, W. F., Rubincam, D. P. \& Burns, J. A. (2000) Dynamical Evolution of Main Belt Meteoroids: Numerical Simulations Incorporating Planetary Perturbations and Yarkovsky Thermal Forces. {\em Icarus, 145}, 301.

\refs Bowell, E., Oszkiewicz, D. A., Wasserman, L. H., Muinonen, K., Penttil\"{a} A., Trilling, D. E. (2014) Asteroid spin-axis longitudes from the Lowell Observatory database. {\em Meteor. \& Plan. Sci., 49}, 95.

\refs Britt, D. T. \& Pieters, C. M., (1991) Black ordinary chondrites: an analysis of abundance and fall frequency. {\em Meteoritics, 26}, 279.

\refs Britt, D. T. \& Pieters, C. M., (1994) Darkening in black and gas-rich ordinary chondrites: The spectral effect of opaque morphology and distribution {\em Geochem. et Cosmoch. Acta, 58}, 3905.

\refs Brown, M. E., Barkume, K. M., Ragozzine, D., Schaller, E. L., (2007) A collisional family of icy objects in the Kuiper belt. {\em Nature, 446}, 294.

\refs Bro\u{z}, M., Morbidelli, A., Bottke, W. F., Rozehnal, J., Vokrouhlick\'{y}, D., Nesvorn\'{y}, D., (2013) Constraining the cometary flux through the asteroid belt during the late heavy bombardment. {\em Astron. \& Astrophys., 551}, A117

\refs Bro\u{z}, \& M., Morbidelli, A., (2013) The Eos Family Halo {\em Icarus, 223}, 844.

\refs Bro\u{z}, \& M., Rozehnal, J., (2011) Eurybates - the only asteroid family among Trojans? {\em Month. Not. Royal. Astron. Soc., 414}, 565.


\refs Burbine, T. H, Gaffey, M. J. \& Bell, J. F., (1992) S-asteroids 387 Aquitania and 980 Anacostia - Possible fragments of the breakup of a spinel-bearing parent body with CO3/CV3 affinities. {\em Meteoritics, 27}, 424.

\refs Burbine, T. \& Binzel, R. P. (2002) Small Main-Belt Asteroid Spectroscopic Survey in the Near-Infrared. {\em Icarus, 159}, 468.

\refs Bus, S.J. \& Binzel, R.P. (2002) Phase II of the Small Main-Belt Asteroid Spectroscopic Survey: A Feature-Based Taxonomy {\em Icarus, 158}, 146.

\refs Campins, H., Hargrove, K., Pinilla-Alonso, N., et al. (2010) Water ice and organics on the surface of the asteroid 24 Themis. {\em Nature, 464}, 1320.

\refs Campins, H., de Le\'{o}n, J., Licandro, J., et al. (2012) Spectra of asteroid families in support of Gaia. {\em Planetary and Space Sci., 73}, 95-97.

\refs Carruba, V., Michtchenko, T. A., Roig, F., Ferraz-Mello, S. \& Nesvorn\'{y}, D. (2005) On the V-type asteroids outside the Vesta family. I. Interplay of nonlinear secular resonances and the Yarkovsky effect: the cases of 956 Elisa and 809 Lundia. {\em Astron. \& Astroph., 441}, 819.

\refs Carruba, V. (2013) An analysis of the Hygiea asteroid family orbital region. {\em Mon. Not. Roy. Astr. Soc., 431}, 3557-3569.

\refs Carruba, V., Domingos, R. C., Nesvorn\'{y}, D., Roig, F., Huaman, M. E., and Souami, D. (2013) A multidomain approach to asteroid families' identification. {\em Mon. Not. Roy. Astr. Soc., 433}, 2075-2096.

\refs Carruba, V., Aljbaae, S., \& Souami, D. (2014) Peculiar Euphrosyne. {\em Astrop. J., 792}, 46.

\refs Carry, B. (2012) Density of asteroids. {\em Planetary \& Space Sci., 73}, 98.

\refs Carvano, J. M., Lazzaro, D., Moth\'{e}-Diniz, T., Angeli, C. A. \& Florczak, M. (2001) Spectroscopic Survey of the Hungaria and Phocaea Dynamical Groups. {\em Icarus, 149}, 173.

\refs Carvano, J. M., Hasselmann, P. H., Lazzaro, D., Moth\'{e}-Diniz, T. (2010) SDSS-based taxonomic classification and orbital distribution of Main Belt asteroids. {\em Astron. \& Astrophys., 510}, A43.

\refs Clark, B. E., Ockert-Bell, M. E., Cloutis, E. A., Nesvorn\'{y}, D., Moth\'{e}-Diniz, T., \& Bus, S.J. (2009) Spectroscopy of K-complex asteroids: Parent bodies of carbonaceous meteorites? {\em Icarus, 202}, 119.

\refs Cellino, A. \& Dell'Oro, A. (2012) The derivation of asteroid physical properties from Gaia observations {\em Plan \& Space Sci., 73}, 52.

\refs Cellino A., Gil-Hutton, R., Tedesco, E.F., Di Martino, M. and Brunini, A. (1999) Polarimetric observations of small observations: preliminary results. {\em Icarus, 138}, 129.

\refs Cellino, A., Zappala, V., Doressoundiram, A. et al. (2001) The Puzzling Case of the Nysa-Polana Family. {\em Icarus, 152}, 225.

\refs Cellino, A., Bus, S.J., Doressoundiram, A. \& Lazzaro, D (2002) Spectroscopic properties of asteroid families. {\em Asteroids III}, 633

\refs Cellino, A., Delb\`{o}, M., Bendjoya, Ph., Tedesco, E. F. (2010) Polarimetric evidence of close similarity between members of the Karin and Koronis dynamical families. {\em Icarus, 209}, 556.

\refs Cellino, A., Dell'Oro, A., Tedesco, E.F. (2009) Asteroid Families: Current Situation. {\em Plan. and Space Sci., 57}, 173.

\refs Cellino, A., Bagnulo, S., Tanga, P., Novakovi\'{c}, B., and Delb\`{o}, M. (2014) A successful search for hidden Barbarians in the Watsonia asteroid family. {\em Mon. Not. Roy. Astr. Soc., 439}, L75-L79.

\refs Christou, A. A. (2013) Orbital clustering of martian Trojans: An asteroid family in the inner Solar system? {\em Icarus, 224}, 144.

\refs Consolmagno, G. J. \& Drake, M. J. (1977) Composition and evolution of the eucrite parent body - Evidence from rare earth elements {\em Geochem. et Cosmochim. Acta, 41}, 1271.

\refs Delb\'{o}, M., Gayon-Markt, J., Busso, G., et al. (2012) Asteroid spectroscopy with Gaia. {\em Plan. \& Space Sci., 73}, 86.

\refs de Leon, J., Licandro, J., Serra-Ricart, M., Pinilla-Alonso, N. \& Campins, H (2010) Observations, compositional, and physical characterization of near-Earth and Mars-crosser asteroids from a spectroscopic survey. {\em Astorn. \& Astroph., 517}, 23.

\refs de Le\'{o}n, J., Pinilla-Alonso, N., Campins, H., Licandro, L., \& Marzo, G. A. (2012) Near-infrared spectroscopic survey of B-type asteroids: Compositional Analysis. {\em Icarus, 218}, 196.

\refs De Luise, F., Dotto, E., Fornasier, S., Barucci, M. A., Pinilla-Alonso, N., Perna, D. \& Marzari, F. (2010) A peculiar family of Jupiter Trojans: The Eurybates. {\em Icarus, 209}.

\refs DeMeo, F. E. \& Carry, B. (2014) Solar system evolution from compositional mapping of the asteroid belt. {\em Nature, 505}, 629.

\refs DeMeo, F. E. \& Carry, B. (2013) The taxonomic distribution of asteroids from multi-filter all-sky photometric surveys. {\em Icarus, 226}, 723.

\refs de Sanctis, M. C., Migliorini, A., Luzia Jasmim, F., Lazzaro, D., et al. (2011) Spectral and mineralogical characterization of inner main-belt V-type asteroids. {\em Astron. \& Astrophys., 533}, A77. 

\refs Dohnanyi, J. S. (1969) Collisional Model of Asteroids and Their Debris. {\em Jour. of Geophys. Research, 74}, 2531.

\refs Doressoundiram, A., Barucci, M.A., Fulchignoni, M. \& Florczak, M. (1998) Eos family: a spectroscopic study. {\em Icarus, 131}, 15. 

\refs Dotto, E., Fornasier, S., Barucci, M. A., et al. (2006) The surface composition of Jupiter Trojans: Visible and near-infrared survey of dynamical families. {\em Icarus, 183}, 420.

\refs Duffard, R. \& Roig, F. (2009) Two new V-type asteroids in the outer Main Belt? {\em Planetary \& Space Sci., 57}, 229.

\refs Dunn, T. L., Burbine, T. H., Bottke, W. F., Clark, J. P. (2013) Mineralogies and source regions of near-Earth asteroids. {\em Icarus, 222}, 273-282.

\refs Durda, D. D., Bottke, W. F., Nesvorn\'{y}, D., et al. (2007) Size-frequency distribution of fragments from SPH/N-body simulations of asteroid impacts: Comparison with observed asteroid families. {\em Icarus, 186}, 498.


\refs Fieber-Beyer, S. K., Gaffey, M. J., Kelley, M. S., Reddy, V., Reynolds, C. M., Hicks, T. (2011) The Maria asteroid family: genetic relationships and a plausible source of mesosiderites near the 3:1 Kirkwood gap. {\em Icarus, 213}, 524.

\refs Fornasier, S., Dotto, E., Marzari, F., et al. (2004) Visible spectroscopic and photometric survey of L5 Trojans: investigation of dynamical families. {\em Icarus, 172}, 221.

\refs Fornasier, S., Dotto, E., Hainaut, O., et al. (2007) Visible spectroscopic and photometric survey of Jupiter Trojans: Final results on dynamical families. {\em Icarus, 190}, 622.

\refs Gaffey, M. J., Burbine, T. H., Piatek, J. L., et al. (1993) Mineralogical variations within the S-type asteroid class. {\em Icarus, 106}, 573.

\refs Gil-Hutton, R., Lazzaro, D. \& Benavidez, P. (2007) Polarimetric observations of Hungaria asteroids. {\em Astron. \& Astrophys., 468}, 1109.

\refs Gladman, B., Michel, P. \& Froeschl\'{e}, C. (2000) The Near-Earth Object Population. {\em Icarus, 146, 176}.

\refs Gomes, R., Levison, H. F., Tsiganis, K., \& Morbidelli, A. (2005) Origin of the cataclysmic Late Heavy Bombardment period of the terrestrial planets {\em Nature, 435}, 26.

\refs Grav, T., Mainzer, A. K., Bauer, J., et al. (2012) WISE/NEOWISE observations of the Hilda population: preliminary results. {\em Astrophys. J., 744}, 197.

\refs Hanu\u{s}, J., Durech, J., Bro\u{z}, M., et al. (2011) A study of asteroid pole-latitude distribution based on an extended set of shape models derived by the lightcurve inversion method. {\em Astron. Astrophys., 530}, A134.

\refs Hanu\u{s}, J., Bro\u{z}, M., Durech, J., et al. (2013) An anisotropic distribution of spin vectors in asteroid families. {\em Astron. Astrophys., 559}, A134.

\refs Harris, A. W, Muller, M., Lisse, C. M. and Cheng, A. F. (2009) A survey of Karin cluster asteroids with the Spitzer Space Telescope. {\em Icarus, 199}, 86.

\refs Hestroffer, D., Dell'Oro, A., Cellino, A. \& Tanga, P. (2010) The Gaia Mission and the Asteroids. {\em Lect. Notes in Phys., 790}, 251.

\refs Hsieh, H. H. \& Jewitt, D. J. (2006) A population of comets in the Main Asteroid Belt. {\em Science, 312}, 561.

\refs Ishihara, D., Onaka, T., Kataza, H., et al. (2010) The AKARI/IRC mid-infrared all-sky survey. {\em Astron. \& Astrophys., 514}, 1.

\refs Ivezi\'{c}, \u{Z}., Lupton, R. H, Juri\'{c}, M. et al. (2002) Color Confirmation of Asteroid Families. {\em Astron. J., 124}, 2943.

\refs Jasmim, F. L., Lazzaro, D., Carvano, J. M. F., Moth\'{e}-Diniz, T., and Hasselmann, P. H. (2013) Mineralogical investigation of several Qp asteroids and their relation to the Vesta family. {\em Astron. Astrophys., 552}, A85.

\refs Jedicke, R., Nesvorn\'{y}, D., Whiteley, R. J., Ivezi\'{c}, \u{Z} and Juri\'{c}, M. (2004) An age-colour relationship for main-belt S-complex asteroids. {\em Nature, 429}, 275.


\refs Jewitt, D., Weaver, H., Agarwal, J., Mutchler, M. \& Drahus, M. (2010) A recent disruption of the main-belt asteroid P/2010A2. {\em Nature, 467}, 817.

\refs Jewitt, D., Weaver, H., Mutchler, M., Larson, S. \& Agarwal, J. (2011) Hubble Space Telescope Observations of Main-belt Comet (596) Scheila. {\em Astrop. J. Lett., 733}, 4.

\refs Kasuga, T., Usui, F., Hasegawa, S., et al. (2012) AKARI/AcuA physical studies of the Cybele asteroid family. {\em Astron. J., 143}, 141.

\refs Kohout, T., Gritsevich, M., Grokhovsky, V. I., et al. (2014) Mineralogy, reflectance spectra, and physical properties of the Chelyabinsk LL5 chondrite - Insight into shock-induced changes in asteroid regoliths. {\em Icarus, 228}, 78.

\refs Kim, M.-J., Choi, Y.-J., Moon, H.-K., et al. (2014) Rotational properties of the Maria asteroid family. {\em Astron. J. 147}, 56.

\refs Kryszczy\'{n}ska, A., Colas, F., Poli\'{n}ska, M., et al. (2012) Do Slivan states exist in the Flora family? I. Photometric survey of the Flora region. {\em Astron. \& Astrophys., 546}, A72.

\refs Lazzaro, D., Moth\'{e}-Diniz, T., Carvano, J. M., et al. (1999) The Eunomia Family: A Visible Spectroscopic Survey {\em Icarus, 142}, 445.

\refs Lazzaro, D., Angeli, C. A., Carvano, J. M., Moth\'{e}-Diniz, T., Duffard, R. \& Florczak, M. (2004) S3OS2: the visible spectroscopic survey of 820 asteroids. {\em Icarus, 172}, 179.

\refs Lebofsky, L. A., Sykes, M. V., Tedesco, E. F., et al. (1986) A refined `standard' thermal model for asteroids based on observations of 1 Ceres and 2 Pallas. {\em Icarus, 68}, 239.

\refs Lebofsky, L. A. \& Spencer, J. R. (1989) Radiometry and a thermal modeling of asteroids. {\em Asteroids II}, 128.

\refs Levison, H. F., Bottke, W. F., Gounelle, M., Morbidelli, A., Nesvorn\'{y}, D., and Tsiganis, K., (2009). Contamination of the asteroid belt by primordial trans-Neptunian objects.  {\em Nature, 460}, 364-366.   	

\refs Licandro, J., Hargrove, K., Kelley, M., et al.(2012) 5-14 $\mu$m Spitzer spectra of Themis family asteroids. {\em Astron. \& Astrophys., 537}, A73.

\refs Mainzer, A. K., Bauer, J., Grav, T., et al. (2011) Preliminary results from NEOWISE: An enhancement to the Wide-field Infrared Survey Explorer for Solar system science {\em Astrophys. J., 731}, 53.

\refs Mainzer, A. K., Bauer, J., Cutri, R.M., et al. (2014) Initial performance of the NEOWISE Reactivation mission {\em Astrophys. J., 792}, 30.


\refs Masiero, J. R., Mainzer, A. K., Grav, T., et al. (2011) Main belt asteroids with WISE/NEOWISE.I. Preliminary albedos and diameters. {\em Astrophys. J., 741}, 68.

\refs Masiero, J. R., Mainzer, A. K., Grav, T., Bauer, J. M. and Jedicke, R. (2012) Revising the age for the Baptistina asteroid family using WISE/NEOWISE data. {\em Astrophys. J., 759}, 14.

\refs Masiero, J. R., Mainzer, A. K., Bauer, J. M., Grav, T., Nugent, C. R. and Stevenson, R. (2013) Asteroid family identification using the hierarchical clustering method and WISE/NEOWISE physical properties. {\em Astrophys. J., 770}, 7.

\refs Masiero, J. R., Grav, T., Mainzer, A. K., et al. (2014) Main Belt Asteroids with WISE/NEOWISE: Near-Infrared Albedos. {\em Astrophys. J., 791}, 121.

\refs Mayne, R. G., Sunshine, J. M., McSween, H. Y., Bus, S. J. \& McCoy, T. J. (2011) The origin of Vesta's crust: Insights from spectroscopy of the Vestoids. {\em Icarus, 214}, 147.

\refs McCord, T. B., Adams, J. B. \& Johnson, T. V. (1970) Asteroid Vesta: Spectral Reflectivity and Compositional Implications. {\em Science, 168}, 1445.


\refs Mignard, F., Cellino, A., Muinonen, K., et al. (2007) The Gaia Mission: Expected Applications to Asteroid Science. {\em Earth, Moon \& Plan., 101}, 97.

\refs Milani, A. (1993) The Trojan asteroid belt: Proper elements, stability, chaos and families. {\em Cel. Mech. and Dynam. Astron., 57}, 59.

\refs Milani, A., Cellino, A., Kne\u{z}evi\'{c}, Z., Novakovi\'{c}, B., Spoto, F. \& Paolicchi, P. (2014) Asteroid families classification: Exploiting very large datasets. {\em Icarus, 239}, 46.

\refs Morbidelli, A., Nesvorn\'{y}, D., Bottke, W.F., Michel, P., Vokrouhlick\'{y}, D. \& Tanga, P. (2003) The shallow magnitude distribution of asteroid families. {\em Icarus, 162}, 328.


\refs Moskovitz, N. A., Willman, M., Burbine, T. H., Binzel, R. P. \& Bus, S. J. (2010) A spectroscopic comparison of HED meteorites and V-type asteroids in the inner Main Belt. {\em Icarus, 208}, 773.

\refs Moskovitz, N. A., Jedicke, R., Gaidos, E., Willman, M., Nesvorn\'{y}, D., Fevig, R. \& Ivezi\'{c}, \u{Z}. (2008) The distribution of balastic asteroids in the Main Belt. {\em Icarus, 198}, 77.

\refs Moth\'{e}-Diniz, T. \& Carvano, J. M. (2005) 221 Eos: a remnant of a partially differentiated parent body? {\em Astron. \& Astroph., 442}, 727.

\refs Moth\'{e}-Diniz, T., Roig, F. and Carvano, J. M. (2005) Reanalysis of asteroid families structure through visible spectroscopy {\em Icarus, 174}, 54.

\refs Moth\'{e}-Diniz, T. Carvano, J. M., Lazzaro, D. (2003) Distribution of taxonomic classes in the main belt asteroids {\em Icarus, 162}, 10.

\refs Moth\'{e}-Diniz, T. Carvano, J. M., Bus, S. J., Duffard, R. \& Burbine, T. H. (2008a) Mineralogical analysis of the Eos family from near-infrared spectra {\em Icarus, 195}, 277.

\refs Moth\'{e}-Diniz, T. \& Nesvorn\'{y}, D. (2008b) Visible spectroscopy of extremely young asteroid families {\em Astron. \& Astrophys. Letters, 486}, 9.

\refs Moth\'{e}-Diniz, T. \& Nesvorn\'{y}, D. (2008c) Tirela: an unusual asteroid family in the outer main belt. {\em Astron. \& Astrophys., 492}, 593.

\refs Murakami, H., Baba, H., Barthel, P., et al., (2007) The Infrared Astronomical Mission AKARI. {\em Pub. Astron. Soc. Japan, 59}, S369.

\refs Nathues, A. (2010) Spectral study of the Eunomia asteroid family Part II: The small bodies {\em Icarus, 208}, 252.

\refs Nathues, A., Mottola, S., Kaasalainen, M., \& Neukum, G. (2005) Spectral study of the Eunomia asteroid family. I. Eunomia. {\em Icarus, 175}, 452.

\refs Neese, C., Ed., Asteroid Taxonomy V6.0. EAR-A-5-DDR-TAXONOMY-V6.0. {\em NASA Planetary Data System}, 2010.

\refs Nesvorn\'{y}, D., Jedicke, R., Whiteley, R.J. and Ivezi\'{c}, \u{Z} (2005) Evidence for asteroid space weathering from the Sloan Digital Sky Survey {\em Icarus, 173}, 132.

\refs Nesvorn\'{y}, D. (2012) Nesvorny HCM Asteroid Families V 2.0. {\em NASA Planetary Data System, EAR-A-VARGBDET-5-NESVORNYFAM-V2.0}.


\refs Novakovi\'{c}, B., Cellino, A., Kne\u{z}evi\'{c}, Z. (2011) Families among high-inclination asteroids. {\em Icarus, 216}, 184.

\refs Oszkiewicz, D. A, Muinonen, K., Bowell, E., et al. (2011) Online multi-parameter phase-curve fitting and application to a large corpus of asteroid photometric data {\em Journ. of Quant. Spectro. and Radiative Trans., 112}, 1919.

\refs Oszkiewicz, D. A, Bowell, E., Wasserman, L. H., Muinonen, K., Penttil\"{a}, A., et al. (2012) Asteroid taxonomic signatures from photometric phase curves. {\em Icarus, 219}, 283.

\refs Parker, A., Ivezi\'{c}, \u{Z}, Juri\'{c}, M., Lupton, R., Sekora, M. D. and Kowalski, A. (2008) The size distributions of asteroid families in the SDSS Moving Object Catalog 4. {\em Icarus, 198}, 138.

\refs Rayner, J. T., Toomey, D. W., Onaka, P. M., et al. (2003) SpeX: A Medium-Resolution 0.8-5.5 Micron Spectrograph and Imager for the NASA Infrared Telescope Facility. {\em Pub. of the Astron. Soc. of the Pacific, 115}, 362.

\refs Reddy, V., Emery, J. P., Gaffey, M. J., Bottke, W. F., Cramer, A. and Kelley, M. S. (2009) Composition of 298 Baptistina: Implications for the K/T impactor link. {\em Meteor. and Plan. Sci., 44}, 1917.

\refs Reddy, V., Carvano, J. M., Lazzaro, D., et al. (2011) Mineralogical characterization of Baptistina asteroid family: Implications for K/T impactor source. {\em Icarus, 216}, 184.

\refs Reddy, V., Sanchez, J. A., Bottke, W. F., et al. (2014) Chelyabinsk meteorite explains unusual spectral properties of Baptistina asteroid family. {\em Icarus, 237}, 116.

\refs Rivkin, A. S. \& Emery, J. P. (2010) Detection of ice and organics on an asteroidal surface. {\em Nature, 464}, 1322.

\refs Roig, F., Nesvorn\'{y}, D., Gil-Hutton, R. \& Lazzaro, D. (2008a) V-type asteroids in the middle main belt. {\em Icarus, 194}, 125.

\refs Roig, F., Ribeiro, A. O. and Gil-Hutton, R. (2008b) Taxonomy of asteroid families among the Jupiter Trojans: comparison between spectroscopic data and the Sloan Digital Sky Survey colors {\em Astron. \& Astrophys., 483}, 911.


\refs Schenk, P., O'Brien, D. P., Marchi, S., et al. (2012) The Geologically Recent Giant Impact Basins at Vesta's South Pole. {\em Science, 336}, 694.

\refs Schunov\'{a}, E., Granvik, M., Jedicke, R., Gronchi, G., Wainscoat, R. \& Abe, S. (2012) Searching for the first near-Earth object family. {\em Icarus, 220}, 1050.


\refs Slivan, S. M., Binzel, R. P., Crespo da Silva, L. D., et al. (2003) Spin vectors in the Koronis family: comprehensive results from two independent analyses of 213 rotation lightcurves. {\em Icarus, 162}, 285.

\refs Slivan, S. M., Binzel, R. P., Kaasalianen, M., et al. (2009) Spin vectors in the Koronis family. II. Additional clustered spins, and one stray. {\em Icarus, 200}, 514.

\refs Slivan, S. M., Binzel, R. P., Boroumand, S. C., et al. (2008) Rotation rates in the Koronis family, complete to H≈11.2 {\em Icarus, 195}, 226.

\refs Solontoi, M. R., Hammergren, M., Gyuk, G. \& Puckett, A. (2012) AVAST survey 0.4-1.0 $\mu$m spectroscopy of igneous asteroids in the inner and middle main belt. {\em Icarus, 220}, 577.

\refs Sunshine, J. M., Bus, S. J., McCoy, T. J., Burbine, T. H., Corrigan, C. M. \& Binzel, R. P. (2004) High-calcium pyroxene as an indicator of igneous differentiation in asteroids and meteorites. {\em Meteor. \& Plan. Sci., 39}, 1343.

\refs Sunshine, J. M., Connolly, H. C., McCoy, T. J., Bus, S. J. \& La Croix, L. M. (2008) Ancient Asteroids Enriched in Refractory Inclusions. {\em Science, 320}, 514.

\refs Szab\'{o}, G. M. and Kiss, L. L. (2008) The shape distribution of asteroid families: Evidence for evolution driven by small impacts {\em Icarus, 196}, 135.

\refs Thomas, C. A., Rivkin, A. S., Trilling, D. E., Enga, M.-T. \& Grier, J. A. (2011) Space weathering of small Koronis family members {\em Icarus, 212}, 158.

\refs Thomas, C. A., Trilling, D. E. \& Rivkin, A. S. (2012) Space weathering of small Koronis family members in the SDSS Moving Object Catalog. {\em Icarus, 219}, 505.

\refs Usui, F., Kuroda, D., M\"{u}ller, T. G., et al. (2011) Asteroid Catalog Using Akari: AKARI/IRC Mid-Infrared Asteroid Survey. {\em Pub. of the Astron. Soc. of Japan, 63}, 1117.

\refs Usui, F., et al. (2015) AKARI infrared spectroscopic observations of asteroids {\em in prep}.

\refs Vernazza, P., Birlan, M., Rossi, A., et al. (2006) Physical characterization of the Karin family. {\em Astron. \& Astrophys, 460}, 945.

\refs Vernazza, P., Binzel, R. P., Thomas, C. A., et al. (2008) Compositional differences between meteorites and near-Earth asteroids {\em Nature, 454}, 858.

\refs Vernazza, P., Binzel, R. P., Rossi, A., Fulchignoni, M. \& Birlan, M. (2009) Solar wind as the origin of rapid reddening of asteroid surfaces {\em Nature, 458}, 993.

\refs Vernazza, P., Zanda, B., Binzel, R. P., et al. (2014) Multiple and Fast: The accretion of ordinary chondrite parent bodies {\em Astroph. J., 791}, 120.

\refs Vokrouhlick\'{y}, D., Nesvorn\'{y}, D. and Bottke, W. F. (2003) The vector alignments of asteroid spins by thermal torques. {\em Nature, 425}, 147.

\refs Walsh, K. J., Delb\'{o}, M., Bottke, W. F., Vokrouhlick\'{y}, D., and Lauretta, D. S. (2013) Introducing the Eulalia and new Polana asteroid families: re-assessing primitive asteroid families in the inner Main Belt. {\em Icarus, 225}, 283-297.

\refs Walsh, K. J., Morbidelli, A., Raymond, S. N., O'Brien, D. P., \& Mandell, A. M. (2012) Populating the asteroid belt from two parent source regions due to the migration of giant planets - ``The Grand Tack''. {\em Meteor. \& Plan. Sci., 47}, 1941.

\refs Warner, B. D., Harris, A. W., Vokrouhlick\'{y}, D., Nesvorn\'{y}, D., Bottke, W. F. (2009) Analysis of the Hungaria asteroid population. {\em Icarus, 204}, 172.

\refs Weaver, H. A., Stern, S. A., Mutchler, M. J., et al. (2006) Discovery of two new satellites of Pluto. {\em Nature, 439}, 943.

\refs Willman, M., Jedicke, R., Nesvorn\'{y}, D., Moskovitz, N., Ivezi\'{c}, \u{Z}., Fevig, R.  (2008) Redetermination of the space weathering rate using spectra of Iannini asteroid family members. {\em Icarus, 195}, 663.

\refs Willman, M., Jedicke, R., Moskovitz, N., Nesvorn\'{y}, D., Vokrouhlick\'{y}, D. \& Moth\'{e}-Diniz, T. (2010) Using the youngest asteroid clusters to constrain the space weathering and gardening rate on S-complex asteroids. {\em Icarus, 208}, 758.

\refs York, D. G., Adelman, J., Anderson, J. E. et al. (2000) The Sloan Digital Sky Survey: Technical Summary. {\em Astron. J., 120}, 1579.

\refs Zappala, V., Cellino, A., Farinella, P. \& Knezevic, Z. (1990) Asteroid families. I - Identification by hierarchical clustering and reliability assessment. {\em Astorn. J., 100}, 2030.

\refs Zappal\`{a}, V., Cellino, A., Dell'Oro, A. \& Paolicchi, P. (2002) Physical and Dynamical Properties of Asteroid Families. {\em Asteroids III}, 619. 

\refs Zellner, B., Tholen, D. J. and Tedesco, E. F., (1985a). The eight-color asteroid survey - results for 589 minor planets. {\em Icarus, 61}, 355.

\refs Zellner, B., Thirunagari, A. and Bender, D., (1985b). The large-scale structure of the asteroid belt. {\em Icarus, 62}, 505-511. 

\refs Ziffer, J., Campins, H., Licandro, J., et al. (2011) Near-infrared spectroscopy of primitive asteroid families {\em Icarus, 213}, 538.

}

\end{document}